\documentclass[12pt]{article} 
\usepackage{graphicx,subfigure,latexsym,amssymb}

\newcommand{\be}{\begin{equation}}
\newcommand{\ee}{\end{equation}}
\newcommand{\bear}{\begin{eqnarray}}
\newcommand{\eear}{\end{eqnarray}}
\newcommand{\ba}{\begin{array}}
\newcommand{\ea}{\end{array}}

\def\({\left(}
\def\){\right)}

\begin{document}

\begin{titlepage}
\vfill
\begin{flushright}
{\normalsize RBRC-1071}\\
\end{flushright}

\vfill
\begin{center}
{\Large\bf Gradient Correction to Photon Emission Rate \\ at Strong Coupling }

\vskip 0.3in
Kiminad A. Mamo$^{1}$\footnote{e-mail: {\tt  kabebe2@uic.edu}} and
Ho-Ung Yee$^{1,2}$\footnote{e-mail:
{\tt hyee@uic.edu}}
\vskip 0.3in

 {\it $^{1}$Department of Physics, University of Illinois,} \\
{\it Chicago, Illinois 60607 }\\[0.15in]
{\it $^{2}$RIKEN-BNL Research Center, Brookhaven National Laboratory,} \\
{\it Upton, New York 11973-5000 }
\\[0.3in]

{\normalsize  2014}

\end{center}

\vfill

\begin{abstract}

We compute the correction to the photon emission rate in first order of shear components of fluid velocity gradients, $\sigma_{ij}$, in near-equilibrium hydrodynamic plasma at strong coupling regime, using the real-time Schwinger-Keldysh formalism in AdS/CFT correspondence.
Our result is an integral of an analytic expression. We observe that the gradient correction to the photon emission rate at strong coupling is about 0.3 - 0.4 times of the equilibrium rate in units of $\sigma_{ij}/T$.

\end{abstract}

\vfill

\end{titlepage}
\setcounter{footnote}{0}

\baselineskip 18pt \pagebreak
\renewcommand{\thepage}{\arabic{page}}
\pagebreak

\section{Introduction}

Electromagnetic probes in heavy-ion collisions are valuable observables that can provide
important information on the properties of quark-gluon plasma. Since the emitted photons and di-leptons rarely interact with the background plasma again, their signals are expected to faithfully describe the state of the quark-gluon plasma at the time of their emissions. More specifically, the photon emission rate with polarization $\epsilon^\mu$ in the static approximation is given by
\be
{d \Gamma\over d^3 \vec k}(\epsilon^\mu)={e^2\over (2\pi)^3 2|\vec k|}\epsilon^\mu (\epsilon^\nu)^* G^<_{\mu\nu}(k)\bigg|_{k^0=|\vec k|}\,,
\label{photonrate}
\ee
where
\be
G^<_{\mu\nu}(k)=\int d^4 x\,e^{-ikx}\langle J_\mu(0) J_\nu(x)\rangle\,,
\ee
so that it can give us information on current correlation functions as well as the temperature of the plasma.
In the static equilibrium plasma, one of the fluctuation-dissipation relations (or KMS relations) can be used to replace the Wightman function $G^<$ with the retarded function $-2 n_B(\omega){\rm Im}G^R$, where $n_B(\omega)$ is the Bose-Einstein distribution.

The perturbative computation of the relevant correlation function assuming a small strong coupling constant $\alpha_s\ll 1$ at sufficiently high temperature has been done for leading order \cite{Kapusta:1991qp,Baier:1991em,Arnold:2001ms} and for the next-leading order \cite{Ghiglieri:2013gia} . Special kinematics such as light-like momenta for photon emission requires to include the infinite set of ladder diagrams that would normally give sub-leading contributions
in ordinary hard thermal loop power counting. 

As there are experimental indications pointing to a strongly coupled nature of quark-gluon plasma created
in heavy-ion collisions, it is important to compute the same in AdS/CFT correspondence to have
the results in other side of the extreme. This has been done in Ref.\cite{CaronHuot:2006te}.

These computations are based on the static equilibrium quark-gluon plasma, and it is clearly desirable to improve the computations for time-dependent, out-of-equilibrium states of quark-gluon plasma.
In general out-of-equilibrium state, the relation $G^<=-2 n_B(\omega){\rm Im}G^R$ no longer holds, which makes the computation much harder since $G^R$ is generally easier to compute (for example, using kinetic theory at weak coupling regime) than $G^<$ in out-of-equilibrium. See Refs.\cite{CaronHuot:2011dr,Chesler:2012zk} which have addressed this issue in some out-of-equilibrium conditions in the AdS/CFT correspondence.
Moreover, the photon emission rate formula (\ref{photonrate}) breaks down if the time scale of photon emission (that is, the formation time of the photon) is larger than the time scale of the evolution of the plasma, and it is applicable only for relatively high frequency photons, with the Fourier transform replaced by the Wigner transform.

A first step to this direction would be to consider the states which are not very far from the equilibrium state, and to compute the required correlation functions in perturbation theory of the deviations from the equilibrium state. Natural such states would be hydrodynamic evolution of near equilibrium plasma where the deviations from equilibrium are organized by derivative expansion. The results should be of high relevance in heavy-ion collisions where a significant portion of plasma evolution is described by hydrodynamics. In Ref.\cite{Dusling:2009bc}, a correction to photon emission rate arising from non-vanishing shear component of velocity gradients (at local rest frame of the fluid)
\be
\sigma_{ij}={1\over 2}\left(\partial_i u_j +\partial_j u_i -{2\over 3}\left(\partial^k u_k\right)\right)\,,
\ee
was computed in the framework of weakly coupled kinetic theory at linear order in $\sigma_{ij}$, which has been further developed and implemented in realistic numerical simulations of heavy-ion collisions in Refs.\cite{Chaudhuri:2011up,Mitra:2011na,Dion:2011pp,Shen:2013cca}.
Considering rotational invariance, the correction to the emission rate at local rest frame should take the following form
\be
{d\Gamma^{\rm shear}\over d^3 \vec k}={e^2 \over T}\Gamma^{(1)}(\omega)\hat k^i\hat k^j \sigma_{ij}\,,
\label{correction}
\ee
where $\hat k^i$ is the unit vector parallel to the momentum direction of the emitted photons, and $T$ is the temperature. In this work, we will compute $\Gamma^{(1)}(\omega)$ in the AdS/CFT correspondence. We note that Ref.\cite{Lekaveckas:2013lha} has computed the similar gradient correction to the drag force on heavy quark, and Ref.\cite{Baier:2012ax}
has computed the photon emission rate in far out-of-equilibrium geometry of falling mass shell \cite{Lin:2006rf}.

The 5 dimensional holographic action we study is
\be
(16\pi G_5){\cal L}_5=R+12-{1\over 8}F_{MN}F^{MN}-{1\over 24\sqrt{6}}{\epsilon^{MNPQR}\over\sqrt{-g_5}}A_M F_{NP}F_{QR}\,,
\ee
with $G_5=\pi/(2N_c^2)$, which describes the U(1) R-symmetry dynamics of $\mathcal{N}=4$ Super Yang-Mills theory of $SU(N_c)$ gauge group \footnote{The strange normalization of the Maxwell term is to be consistent with the previous choice in Ref.\cite{CaronHuot:2006te}. This does not affect the ratio between the zero'th and the first order correction shown in Figure \ref{fig10}.}. The last Chern-Simons term is a holographic manifestation of $U(1)_R^3$ triangle anomaly, but is irrelevant for our subsequent discussion. Although this theory is not precisely QCD per se, it is a useful benchmark theory to understand strongly coupled gauge theory in general. We will discuss later how one may reasonably translate the results in this theory to those in QCD.

Our end result seems quite compact: it is given by
\bear
\Gamma^{(1)}(\omega)&=&{1\over (2\pi)^3 2\omega} 2n_B(\omega){\frac{N_c^2 T^2}{16\pi}}\label{intro1}\\ &\times& {\rm Im}\left[{1\over C^2}\int_0^1 du\,\Bigg(f(u) (\partial_u \bar S(u))H(u)(\partial_u H(u))+\left(\omega\over 2\pi T\right)^2 {1\over u}\,\bar S(u) H(u)^2\Bigg)\right]\,,\nonumber
\eear
where $n_B(\omega)=1/(e^{\omega/T}-1)$ is the Bose-Einstein distribution, and
\bear
&&f(u)=1-u^2\,,\quad \bar S(u)=\pi-2\arctan\left(1\over \sqrt{u}\right)+\log\left((1+\sqrt{u})^2(1+u)\right)\,,\\
&& H(u)=(1-u)^{-i{\omega\over 4\pi T}}(1+u)^{-{\omega\over 2\pi T}}{_2F_1}\left(1-{1\over 2}(1+i){\omega\over 2\pi T},-{1\over 2}(1+i){\omega\over 2\pi T};1-i{\omega\over 2\pi T};{1\over 2}(1-u)\right),\nonumber\\
&& C={_2F_1}\left(1-{1\over 2}(1+i){\omega\over 2\pi T},-{1\over 2}(1+i){\omega\over 2\pi T};1-i{\omega\over 2\pi T};{1\over 2}\right)\,.\nonumber
\eear
Although we need to perform the integration numerically, it is remarkable and useful to have the above analytic expression.

\section{Schwinger-Keldysh formalism in AdS/CFT correspondence }

\begin{figure}[t]
	\centering
	\includegraphics[width=9cm]{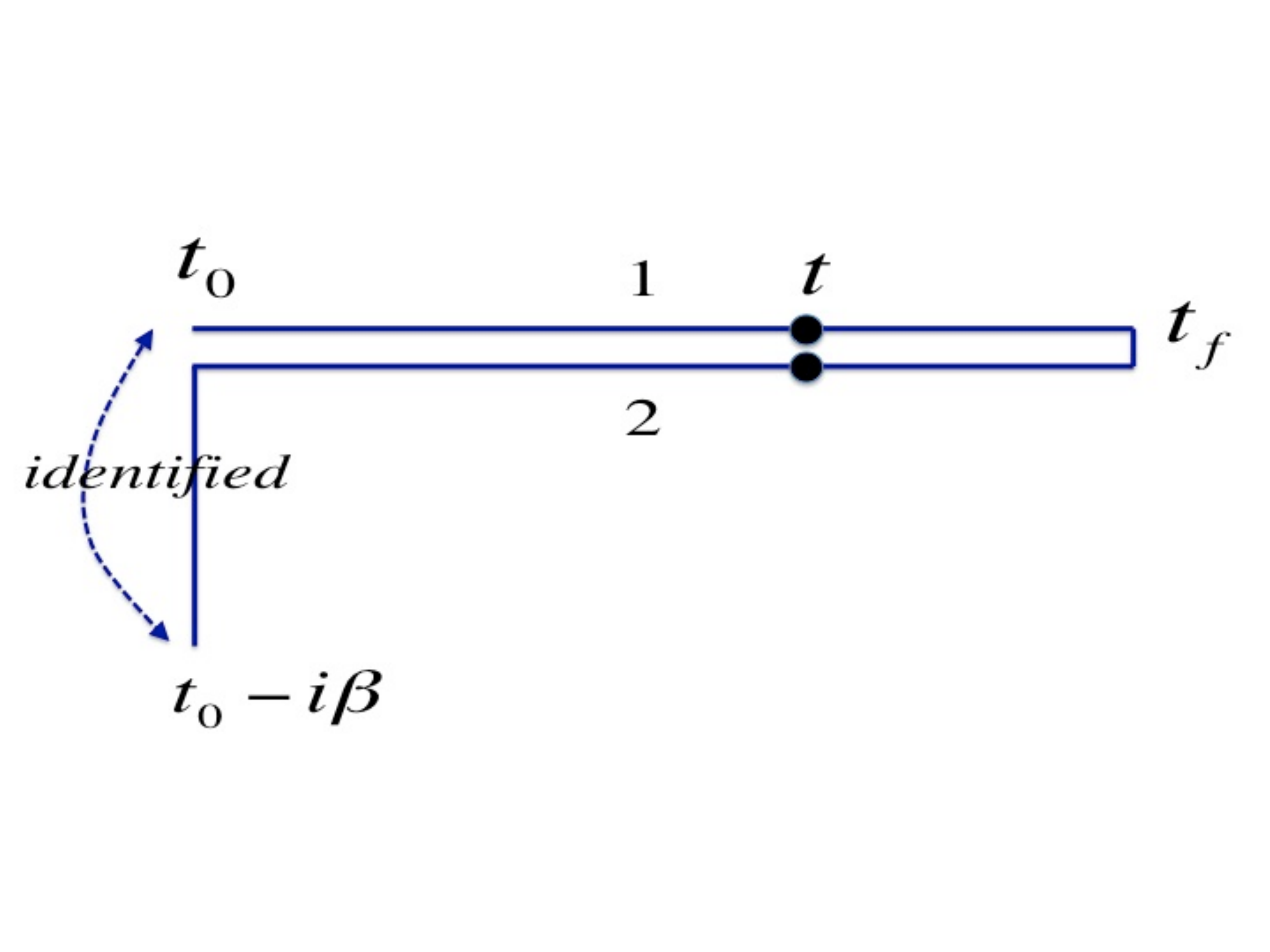}
		\caption{Schwinger-Keldysh contour.\label{fig1}}
\end{figure}

We briefly review the Schwinger-Keldysh formalism in the AdS/CFT correspondence~\cite{Herzog:2002pc,Barnes:2010jp,Skenderis:2008dg}.
The Schwinger-Keldysh formalism is a nice way of describing various real-time correlation functions with different time orderings. In the field theory side, it is a field theory defined on a complex time contour shown in Figure \ref{fig1}: the line labeled by 1 describes a unitary time evolution of the state ensemble prepared by the thermal imaginary time contour situated at the far left, while the line 2 describes the time evolution of the conjugate state, that is, the bra states. Because of this difference given by complex conjugation, the path integral measure for line 2 is $e^{-iS_2}$, not $e^{iS_2}$. One can also think of this as time flowing from future to past.
Therefore, the resulting path integral of any operator insertion gives us the real-time expectation value of that operator sandwiched between bra and ket states with thermal ensemble: this is precisely what we would like to compute. It is very important to have the boundary condition at $t=t_f$ such that the field variables from line 1 and those from line 2 are equal at that point.

By putting operators in suitable positions in the Schwinger-Keldysh contour, one can achieve any operator ordering. Our needed Wightman function can be obtained for example by
\be
G^<_{\mu\nu}(x)\equiv \langle \hat J_\mu(0) \hat J_\nu(x)\rangle=\langle J^{(2)}_\mu(0) J^{(1)}_\nu(x)\rangle_{\rm SK}\,,
\ee
where $\hat J$ means an operator acting on the Hilbert space, while $J^{(1,2)}$ mean the field variables on the lines 1 and 2 in the Schwinger-Keldysh path integral.

It is convenient to introduce ``ra''-basis for any (composite) variable generically denoted by $\phi$ as
\be
\phi^{(r)}={1\over 2}\left(\phi^{(1)}+\phi^{(2)}\right)\,,\quad
\phi^{(a)}=\phi^{(1)}-\phi^{(2)}\,.
\ee
The boundary condition at $t=t_f$ is translated to the condition $\phi^{(a)}(t_f)=0$, and it is generally true that any correlation functions with an $(a)$-type object having the latest time always vanishes. In terms of (ra) variables, we have
\bear
G^<_{\mu\nu}(x)&=&\langle J^{(2)}_\mu(0) J^{(1)}_\nu(x)\rangle_{\rm SK}\nonumber\\&=&\langle J^{(r)}_\mu(0) J^{(r)}_\nu(x)\rangle_{\rm SK}-{1\over 2}\langle J^{(a)}_\mu(0) J^{(r)}_\nu(x)\rangle_{\rm SK}+{1\over 2}\langle J^{(r)}_\mu(0) J^{(a)}_\nu(x)\rangle_{\rm SK}\nonumber\\&\equiv& G^{(rr)}_{\nu\mu}(x)-{1\over 2}G^{(ra)}_{\nu\mu}(x)+{1\over 2}G^{(ar)}_{\nu\mu}(x) \,,\label{grrraar}
\eear
so the computation of $G^<$ necessitates the computation of $G^{(rr)}$, $G^{(ra)}$ and $G^{(ar)}$.
Going back to the operator formalism, one can easily check that $G^{(ra)}$ is proportional to the retarded correlation function, and $G^{(ar)}$ the advanced correlation function. $G^{(rr)}$ generally encodes fluctuations provided by both quantum noise and occupation numbers. The subtraction of $(1/2)(G^{(ra)}-G^{(ar)})$ in the above expression removes the quantum noise, so that $G^<$ for positive frequency $\omega$ encodes contributions only from the occupation numbers, and it is understandable why $G^<$ gives the emission rates originating from the occupied states. In equilibrium, the relation $G^<=-2n_B(\omega){\rm Im}G^R$ illustrates this point, where $-{\rm Im}G^R$ is the density of states with frequency $\omega$.

The free theory Schwinger-Keldysh path integral, including the imaginary time thermal circle, is Gaussian, so that any perturbation theory from the free limit admits the Feynman diagram expansion. This allows one to compute any real-time correlation functions in Feynman diagram expansion, which is a great theoretical convenience.

\begin{figure}[t]
	\centering
	\includegraphics[width=9cm]{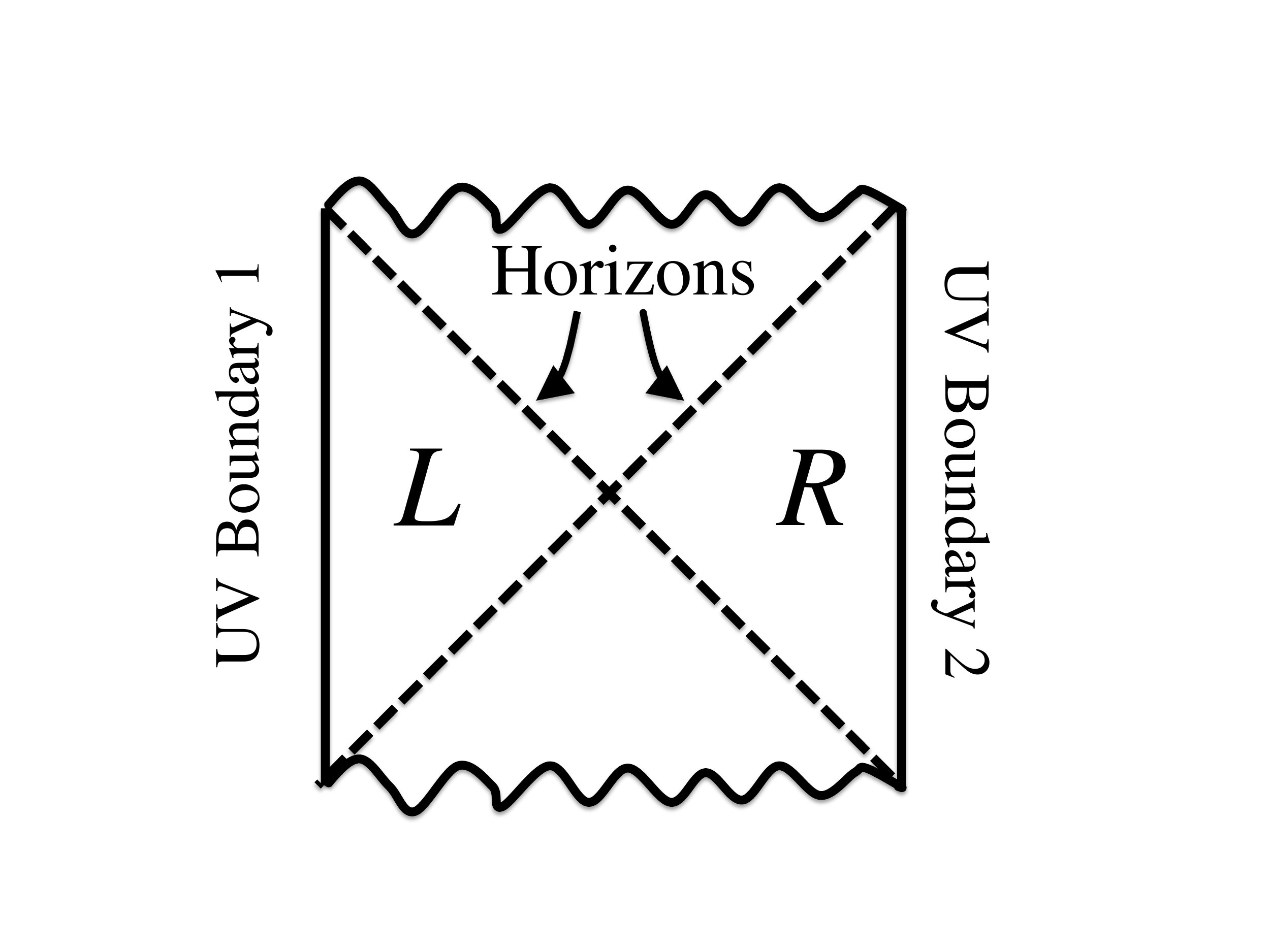}
		\caption{Penrose diagram of AdS black-hole geometry.\label{fig2}}
\end{figure}

So far, we have reviewed the field theory side of Schwinger-Keldysh formalism. In thermal AdS black-hole geometry, there is an intuitive correspondence between the geometry and the Schwinger-Keldysh formalism, as illustrated in Figure \ref{fig2} \cite{Israel:1976ur,Maldacena:2001kr}. The analytically continued AdS black-hole geometry consists of four parts, and the two parts labels by L (left) and R (right) have UV boundaries that can be identified with the lines 1 and 2 in the field theory Schwinger-Keldysh formalism. Indeed the natural time direction from the analytic property of the geometry is reversed from L to R, reminiscent of the similar concept in Schwinger-Keldysh formalism.

The holographic bulk actions and the fields in L and R regions can be thought of as holographic descriptions of field theory parts living on lines 1 and 2 respectively \cite{Herzog:2002pc,Barnes:2010jp}. Correspondingly, the bulk action in the R region appears in the bulk path integral as $e^{-iS_R}$ contrary to that for the L region, $e^{iS_L}$.
One also naturally introduces the same ``ra'' variables for the bulk fields. The UV boundary values of the bulk (ra) fields correspond to the sources of (ra)-type in the field theory, which couple to (ar)-type operators
in the field theory according to the AdS/CFT correspondence. By performing variations of the bulk path integral with respect to these UV boundary values of (ra)-type fields, one can obtain real-time correlation functions of any kind.
 The meaning of the regions inside the horizons is yet unclear to our current understanding, but one can forgo them by specifying suitable boundary conditions on the horizons~\cite{Herzog:2002pc} and dealing with only L and R regions outside. Within this set-up, the correct KMS relations among real-time correlation functions have been checked to hold true for the free quadratic bulk theory in equilibrium black-hole geometry, corresponding to equilibrium finite temperature plasma in the gauge theory side \cite{Herzog:2002pc,Barnes:2010jp,Skenderis:2008dg}.

 The leading bulk holographic action is typically quadratic in fields, that is, Gaussian, and higher order interactions can be treated perturbatively.
 Although this real-time perturbation theory in the bulk looks technically quite similar to the real-time perturbation theory in the gauge theory side, the bulk perturbative expansion maps to the large $N_c$ expansion in the gauge theory side. At the quadratic leading level, the bulk theory defined by a path integral of the Gaussian action, with a chosen Dirichlet boundary condition at the UV boundary and the suitable horizon boundary condition, is a 5 dimensional free quantum field theory on its own, living in a curved geometry. According to AdS/CFT correspondence, the mapping to the gauge theory side is achieved by the dependence of this 5 dimensional quantum field theory on the specific UV Dirichlet boundary values of the bulk fields, which act as external sources coupled to gauge theory operators.
By variations of these UV boundary conditions, one can thus obtain correlation functions of gauge theory operators.

Since the bulk theory with a fixed boundary condition is a quantum field theory (of bulk fields) defined on a curved AdS geometry, one can consider various types of real-time correlation functions defined in the bulk space-time. For example, in our case of free U(1) Maxwell theory in 5 dimensions, the real-time two point functions are
\be
g^{(\alpha\beta)}_{\mu\nu}(x,r|x',r')\equiv \langle A^{(\alpha)}_\mu(x,r) A^{(\beta)}_\nu(x',r')\rangle_{\rm 5D}\,,\label{bulktobulk}
\ee
where $\alpha, \beta$ are either $1, 2$, or $r, a$ depending on which type of ordering one would like to compute in Schwinger-Keldysh formalism.
These objects are previously called bulk-to-bulk propagators \cite{CaronHuot:2011dr}. Note that they are observables in the 5 dimensional quantum theory with a fixed UV boundary condition (Dirichlet), and {\it a priori} they are quite different objects than the gauge theory correlation functions obtained by {\it variations} of UV boundary conditions. However, we will shortly see that it is possible to read off gauge theory correlation functions from the bulk-to-bulk propagators (see subsection \ref{subsec}).
This relation, previously found in Ref.\cite{CaronHuot:2011dr}, has been referred to as the relations between bulk-to-bulk, boundary-to-bulk, and boundary-to-boundary propagators. Conceptually, the idea is similar to that of holographic renormalization where one can read off gauge theory one-point function from the bulk field profile \cite{de Haro:2000xn}. In subsection \ref{subsec}, we will clarify that this relation is a generic consequence of any Gaussian theory, although it may also be valid beyond Gaussian limit.
Having the above mentioned relation greatly simplifies the computations of gauge theory real-time correlation functions via AdS/CFT correspondence: one only needs to compute the bulk real-time correlation functions of the bulk quantum theory with a {\it fixed} UV Dirichlet boundary condition,
without considering variations of the UV boundary condition.
The computation of any type of bulk-to-bulk correlation functions can be performed in the framework of Schwinger-Keldysh real-time perturbation theory as discussed before~\cite{Barnes:2010jp,Arnold:2010ir}. Once this is achieved, one can explore the relation between bulk-to-bulk and boundary-to-boundary correlation functions to finally get the gauge theory correlation functions.

\begin{figure}[t]
	\centering
	\includegraphics[width=8cm]{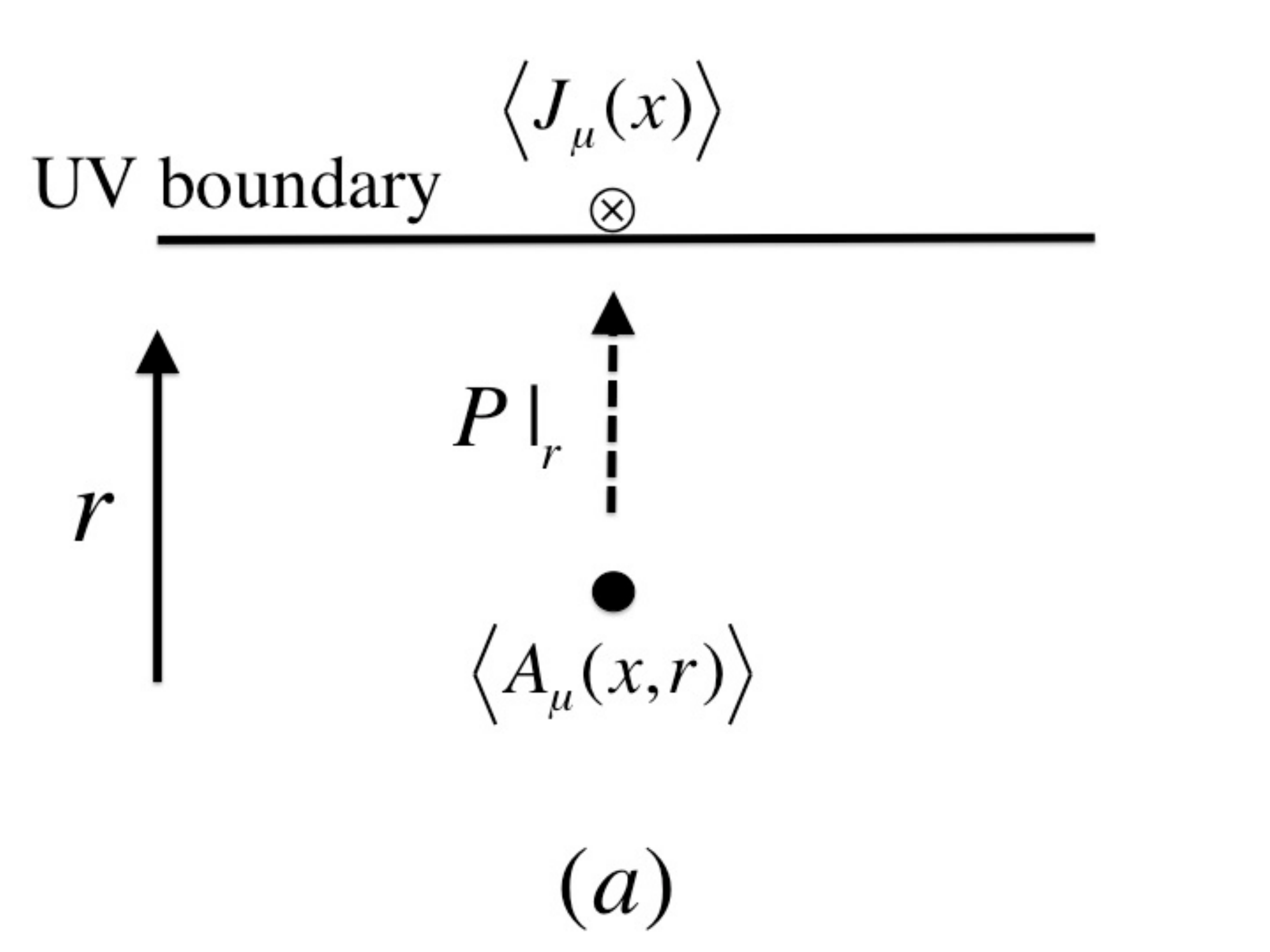}\includegraphics[width=8cm]{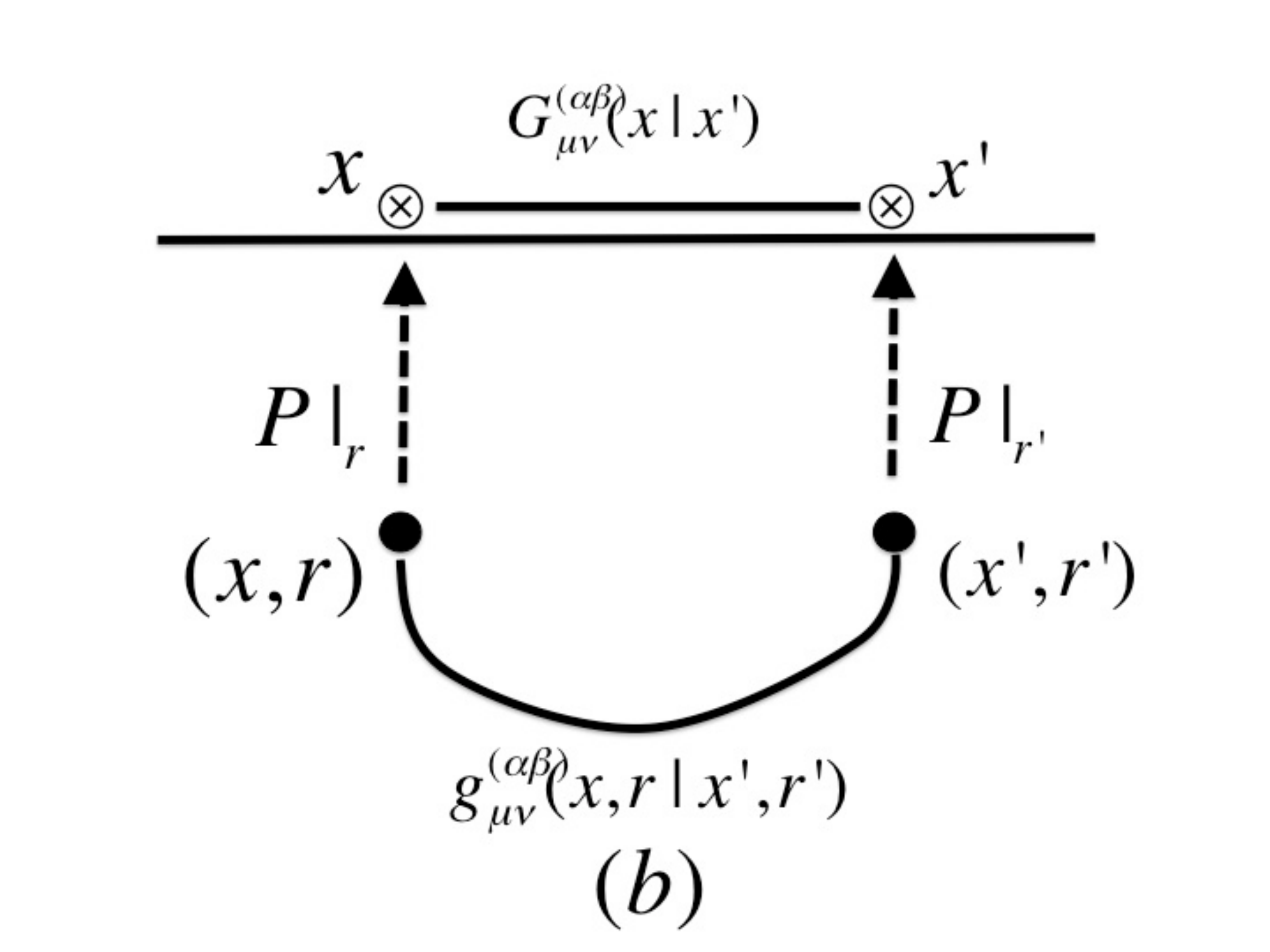}
		\caption{(a) Bulk to boundary relation for 1-point functions. (b) Bulk to boundary relation for 2-point functions. It should generalize to $n$-point functions as well. \label{fig2x}}
\end{figure}

Let us recall that in the asymptotic AdS geometry near UV boundary $r\to\infty$,
\be
ds^2={dr^2\over r^2}+r^2\eta_{\mu\nu}dx^\mu dx^\nu\,,\quad \eta={\rm diag}(-+++)\,,
\ee
the solution of Maxwell's equation in 5 dimensions takes the form in $A_r=0$ gauge
\be
A_\mu(x,r)\sim A^{(0)}_\mu(x) +{\tilde A_\mu(x)\over r^2}+{-2 B_\mu(x)\log r\over r^2}+\cdots\,,
\ee
where $\tilde A_\mu(x)$ is an arbitrary source which couples to the gauge theory U(1) current $J^\mu(x)$, and we have
\be
B_\mu(x)={1\over 4}\partial^\nu F^{(0)}_{\mu\nu}(x)\,.
\ee
The $\tilde A_\mu(x)$ is determined only by IR dynamics. According to holographic renormalization \cite{de Haro:2000xn,Sahoo:2010sp}, the gauge theory current expectation value is given by
\be
\langle J_\mu(x)\rangle={1\over 16\pi G_5}\left(\tilde A_\mu(x)+B_\mu(x)\right)\,,\label{onep1}
\ee
where the last term is ambiguous up to finite counter-term, and we can and will drop it.
This result motivates the following definition of a projection operator ${\cal P}$,
\be
{\cal P}|_r\cdot A_\mu(r,x)\equiv {1\over 16\pi G_5} \tilde A_\mu(x)\,.\label{onep2}
\ee
Note that this is in general {\it not} equal to
\be
-{1\over 32\pi G_5} \lim_{r\to\infty}  {r^3}\partial_r A_\mu(x,r)\,,
\ee
due to the presence of $B_\mu(x)$ term. The ${\cal P}$ gives the relation between the bulk one point function (that is, the profile $A_\mu(x,r)$) and the gauge theory one point function $\langle J_\mu(x)\rangle$, as illustrated in Figure \ref{fig2x}(a).

\begin{figure}[t]
	\centering
	\includegraphics[width=8cm]{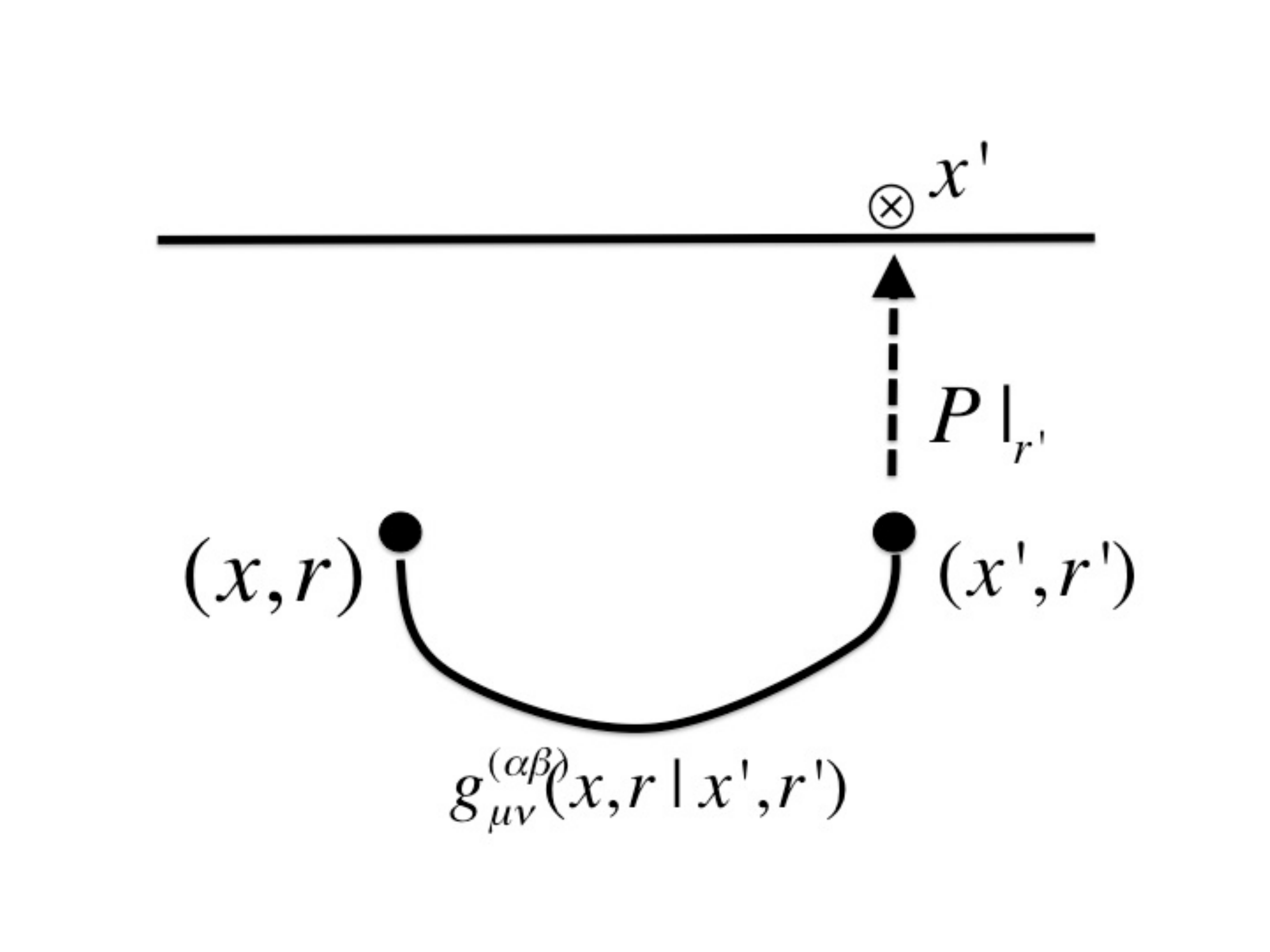}\includegraphics[width=8cm]{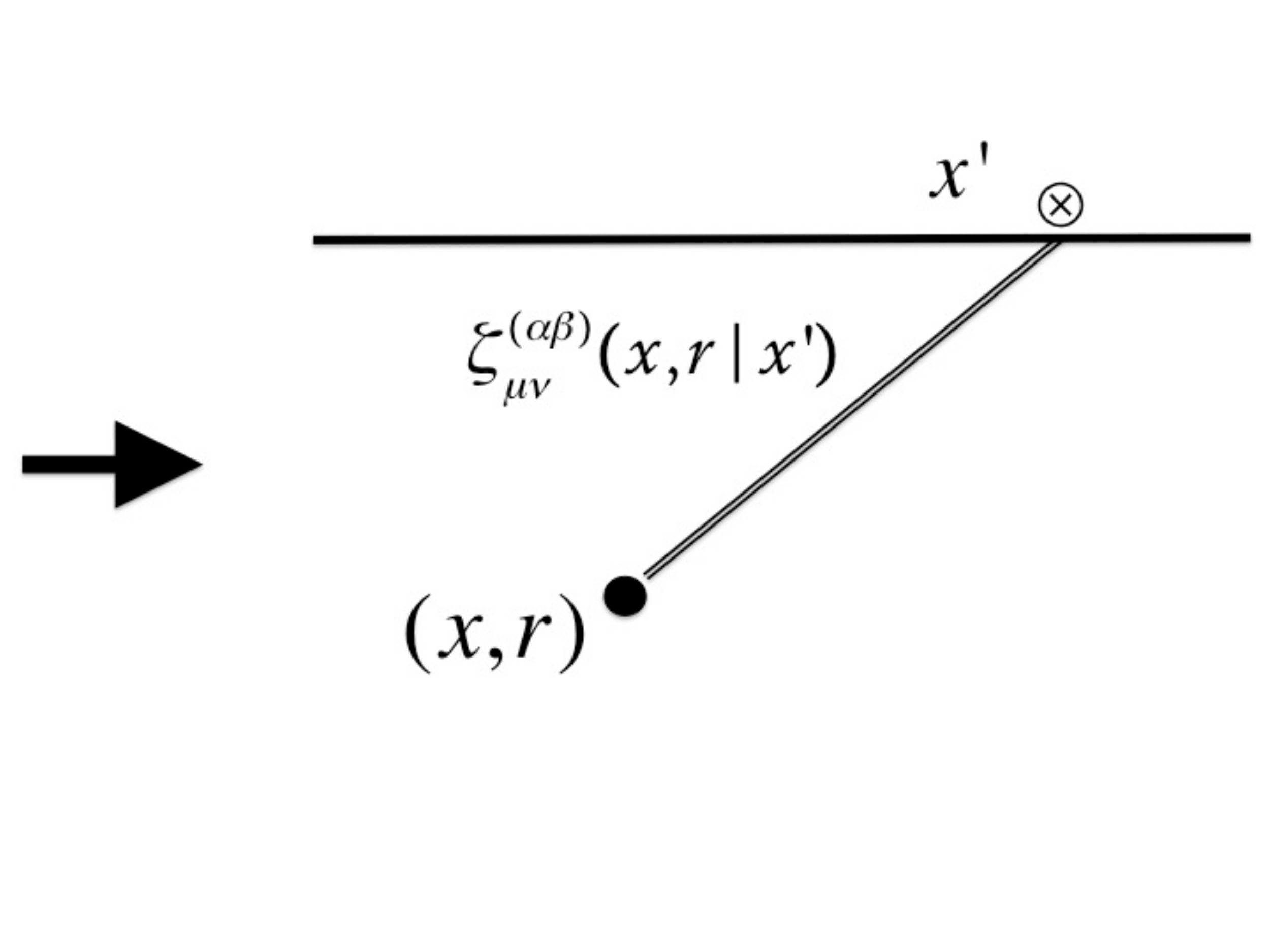}
		\caption{Boundary-to-bulk propagator ${\cal G}^{(\alpha\beta)}_{\mu\nu}(x,r|x')$ (the doubled line on the right).\label{fig3}}
\end{figure}

Let us define real-time gauge theory two point functions as
\be
G^{(\alpha\beta)}_{\mu\nu}(x|x')\equiv \langle J_\mu^{(\alpha)}(x)J_\nu^{(\beta)}(x')\rangle_{\rm 4D}\,,\label{bdtobd}
\ee
where $\alpha,\beta$ are $1,2$ or $r,a$ types, which are also called boundary-to-boundary propagators. They are what we would like to compute at the end.
The claim is that they can be computed from the bulk-to-bulk propagators (\ref{bulktobulk}) by
\be
G^{(\alpha\beta)}_{\mu\nu}(x|x')={\cal P}|_r {\cal P}|_{r'}\cdot
g^{(\alpha\beta)}_{\mu\nu}(x,r|x',r')\,,\label{projection}
\ee
which is a simple generalization of (\ref{onep1}) and (\ref{onep2}) for the one point functions. We will illustrate this in subsection \ref{subsec}. Figure \ref{fig2x}(b) depicts this relation pictorially.
It will be useful to have one more object, boundary-to-bulk propagator, defined by
\be
{\cal G}^{(\alpha\beta)}_{\mu\nu}(x,r|x')\equiv {\cal P}|_{r'}\cdot g^{(\alpha\beta)}_{\mu\nu}(x,r|x',r')\,,
\label{boundarytobulk}\ee
so that
\be
G^{(\alpha\beta)}_{\mu\nu}(x|x')={\cal P}|_r \cdot {\cal G}^{(\alpha\beta)}_{\mu\nu}(x,r|x')\,.\label{bdfrombulk}
\ee See Figure \ref{fig3} for pictorial representation of boundary-to-bulk propagators ${\cal G}^{(\alpha\beta)}_{\mu\nu}(x,r|x')$ (the doubled line on the right).
We now discuss several salient properties of these objects.

1) Since the bulk-to-bulk propagators (\ref{bulktobulk}),
\be
g^{(\alpha\beta)}_{\mu\nu}(x,r|x',r')\equiv \langle A^{(\alpha)}_\mu(x,r) A^{(\beta)}_\nu(x',r')\rangle_{\rm 5D}\,,\label{bulktobulk2}
\ee
are correlation functions of the bulk quantum theory with the Dirichlet UV boundary condition,
they should vanish when either $r$ or $r'$ goes to the UV boundary,
\be
\lim_{r\to\infty}g^{(\alpha\beta)}_{\mu\nu}(x,r|x',r')=\lim_{r'\to\infty}g^{(\alpha\beta)}_{\mu\nu}(x,r|x',r')=0\,.
\ee

2) In the case of leading Gaussian theory in the bulk (as in our case of free Maxwell theory),
the bulk-to-bulk propagators $g^{(\alpha\beta)}_{\mu\nu}(x,r|x',r')$ are naturally the Green's functions of the bulk kinetic term, so that they satisfy
\bear
D^M\langle F_{M\mu}^{(\alpha)}(x,r) A^{(\beta)}_\nu(x',r')\rangle&=&D^MD_M g^{(\alpha\beta)}_{\mu\nu}(x,r|x',r')-D^MD_\mu g^{(\alpha\beta)}_{M\nu}(x,r|x',r')\nonumber\\
&=&i(32\pi G_5)\eta_{\mu\nu}\eta^{\alpha\beta}{1\over\sqrt{-g_5}}\delta^{(5)}(x,r|x',r')\,,\label{bulkeq}
\eear
where the metric covariant derivative $D^M$ acts on the $(x,r)$ indices, and the $\eta^{\alpha\beta}$ symbol is given by
\be
\eta^{\alpha\beta}=\left(\begin{array}{cc} 1&0\\0&-1\end{array}\right)
\ee
in the (1,2) basis, or
\be
\eta^{\alpha\beta}=\left(\begin{array}{cc} 0&1\\1&0\end{array}\right)
\ee
in the (r,a) basis.
The similar equation also applies to the $(x',r')$ indices. These equations can determine $g^{(\alpha\beta)}_{\mu\nu}(x,r|x',r')$ completely, given the UV boundary condition 1) and the appropriate horizon boundary conditions. More precisely, one has to fix the gauge to find a well-defined $g^{(\alpha\beta)}_{\mu\nu}(x,r|x',r')$: one possibility is to introduce $R_\xi$-gauge fixing term $-{1\over 2\xi}\left(D^M A_M\right)^2$ in the action. However, in our computation of photon emission rate, we will contract the vector indices with the transverse photon polarization tensor $\epsilon^\mu$ with $\epsilon^0=0$ and $\vec\epsilon\cdot\vec k=\epsilon^i k_i=0$ in the Fourier space, and the resulting transverse part is physical and unambiguously given independent of gauge choice. In our work, we will assume $A_r=0$ gauge. Therefore, we will not discuss about the detailed from of $g^{(\alpha\beta)}_{\mu\nu}(x,r|x',r')$, except its unambiguous polarization-contracted form that we present in section \ref{mainsec}.

3) Since the projection operator $\cal P$ is linear, the same type of equation without the right-hand side is satisfied by the bulk-to-boundary propagators,
\be
D^MD_M {\cal G}^{(\alpha\beta)}_{\mu\nu}(x,r|x')-D^MD_\mu {\cal G}^{(\alpha\beta)}_{M\nu}(x,r|x')=0\,.\label{bulkeom2}
\ee
What is somewhat non-trivial to see is that the UV boundary condition for
${\cal G}^{(\alpha\beta)}_{\mu\nu}(x,r|x')$ is given by
\be
\lim_{r\to\infty} {\cal G}^{(\alpha\beta)}_{\mu\nu}(x,r|x')=(-i)\eta_{\mu\nu}\eta^{\alpha\beta} \delta^{(4)}(x-x')\,.\label{UVboundary}
\ee
This relation was first derived for a free scalar field in Ref.\cite{CaronHuot:2011dr},
and following the similar steps one can derive it from the equation (\ref{bulkeq}) above. In subsection \ref{subsec}, we will see that this is a generic consequence of any Gaussian theory with Dirichlet boundary condition.
With the horizon boundary condition, the above equation and the UV boundary condition uniquely fix the boundary-to-bulk propagators ${\cal G}^{(\alpha\beta)}_{\mu\nu}(x,r|x')$ in the equilibrium black-hole background.

4) It has been known that a free field theory living in a black-hole background features a finite temperature ensemble with the Hawking temperature. In the case of AdS where there exists a UV boundary, the equilibrium quantum state can be achieved which defines an appropriate stationary state, and the horizon boundary conditions follow from it.
Therefore, it is natural to have the KMS relations satisfied with the bulk-to-bulk propagators $g^{(\alpha\beta)}_{\mu\nu}(x,r|x',r')$ with the Hawking temperature. This has been referred to as bulk fluctuation-dissipation relations in literature \cite{deBoer:2008gu,Son:2009vu}. Since the projection operator $\cal P$ is linear, the same KMS relation should also hold for the gauge theory correlation functions $G^{(\alpha\beta)}_{\mu\nu}(x|x')$ which are obtained via (\ref{projection}), which proves the gauge theory KMS relations with the same Hawking temperature in equilibrium.

5) By definition, it is trivial to see
\be
g^{(ra)}_{\mu\nu}(x,r|x',r')=g^{(ar)}_{\nu\mu}(x',r'|x,r)\,.\label{ravsar}
\ee
What is {\it not} trivial is the identity
\be
{\cal G}^{(ra)}_{\mu\nu}(x,r|x')={\cal G}^{(ar)}_{\nu\mu}(x',r|x)\,.\label{ravsar2}
\ee
Note that one can't derive (\ref{ravsar2}) from (\ref{ravsar}). To prove this, first observe that ${\cal G}^{(ra)}_{\mu\nu}(x,r|x')$ and ${\cal G}^{(ar)}_{\nu\mu}(x',r|x)$ satisfy the same UV boundary condition via (\ref{UVboundary}),
\be
\lim_{r\to\infty}{\cal G}^{(ra)}_{\mu\nu}(x,r|x')=\lim_{r\to\infty}{\cal G}^{(ar)}_{\nu\mu}(x',r|x)=(-i)\eta_{\mu\nu}\delta^{(4)}(x-x')\,.
\ee
From the relation (\ref{bdfrombulk}) between the gauge theory correlation functions and the boundary-to-bulk propagators, one sees that
\be
{\cal P}\cdot {\cal G}^{(ra)}_{\mu\nu}(x,r|x')=G^{(ra)}_{\mu\nu}(x,x')\,,\quad {\cal P}\cdot {\cal G}^{(ar)}_{\nu\mu}(x',r|x)=G^{(ar)}_{\nu\mu}(x',x)\,.
\ee
On the other hand, in the gauge theory it is clear that $G^{(ra)}_{\mu\nu}(x,x')=G^{(ar)}_{\nu\mu}(x',x)$ by definition. Noting that the projection operator $\cal P$ defined in (\ref{onep2}) picks up the sub-leading normalizable mode of the $r$ dependence, we see that ${\cal G}^{(ra)}_{\mu\nu}(x,r|x')$ and ${\cal G}^{(ar)}_{\nu\mu}(x',r|x)$ have the same UV boundary values and the normalizable modes. Since they both satisfy the second order Maxwell equation, we conclude that they must be identical.

In the next subsection, we will illustrate the generality of some of the above mentioned properties in a simple discrete toy ``XY'' model.

\subsection{Toy ``XY'' model\label{subsec}}

In this subsection, we would like to illustrate some of the mentioned properties above in the case of simple Gaussian path integral of finite number of degrees of freedom, to show that these properties are in fact general features of any Gaussian theory in the holographic bulk in AdS/CFT correspondence.

In the AdS/CFT correspondence, the bulk fields consist of two independent pieces: the non-normalizable modes which can be identified as the (properly scaled) values of the bulk fields at the UV boundary, and the normalizable modes which are dynamical degrees of freedom in the bulk. In our simple toy model, let us call the non-normalizable modes (or the UV boundary values) $X^a$ collectively with alphabetical indices, and the normalizable dynamical degrees of freedom $Y^I$ with capital indices.
The action of our toy ``XY'' model is a Gaussian one
\be
S(X,Y)=Y^I F_{Ia} X^a+ {1\over 2}Y^I M_{IJ} Y^J =Y^T F X+{1\over 2} Y^T M Y\,,
\ee
with some constant matrices $F_{Ia}$ and $M_{IJ}$, and the last expression is a matrix notation with $T$ meaning transpose. The first term represents the fact that the UV boundary value of bulk fields are related to the dynamical bulk profile of the fields by the UV asymptotic kinetic term in the action, and the second term is simply the dynamical kinetic term for the normalizable modes.

The above description is valid even for our real-time AdS/CFT correspondence with AdS black-hole geometry: we include only the bulk fields inside the L and R regions of the Penrose diagram of Figure \ref{fig2} (the $a$ or $I$ indices include both L and R regions), while we can neglect the regions beyond event horizon by simply fixing suitable boundary conditions on our fields on the horizon. Since the horizon boundary condition is a linear relation imposed on the field profile, restricting the field profiles to those satisfying the specified horizon boundary conditions defines a linear subspace in the vector space of field profiles. One can simply redefine the variables $Y^I$ such that they now represent the field profiles living in this subspace of fields satisfying horizon boundary conditions, and the resulting action for the redefined $Y^I$ is clearly Gaussian again due to the linearity of the horizon boundary condition imposed.

In the AdS/CFT correspondence, one performs the path integral of the above action with {\it fixed} values of $X^a$;
\be
\int {\cal D}Y^I\,\,e^{iS(X,Y)}\equiv e^{iW(X)}\,,
\ee
to define a boundary effective action $W(X)$. The $X^a$ is interpreted as a source coupled to a boundary operator $J_a$ in the boundary field theory, so that
one and two-point functions of the boundary operator $J_a$ in the boundary field theory can be obtained from $W(X)$ by (the extra $(-i)$ for $G_{ab}$ is due to the $i$ in the path integral measure of $iS$)
\be
\langle J_a\rangle={\delta W(X)\over \delta X^a}\,,\quad G_{ab}\equiv \langle J_a J_b\rangle_{\rm connected} =(-i) {\delta^2 W(X)\over\delta X^a\delta X^b}\,.\label{onetwo}
\ee
Since the action is Gaussian, one can easily find the stationary point of $Y$ from the equation of motion,
\be
FX+MY^{\rm on-shell}=0\quad\longrightarrow\quad Y^{\rm on-shell}=-M^{-1} F X\,,\label{onshell}
\ee
and the effective action $W(X)$ is simply given by the action $S$ evaluated with the solution of the equation of motion, $Y^{\rm on-shell}$,
\be
W(X)=S(X,Y^{\rm on-shell})=\left(Y^T F X+{1\over 2} Y^T M Y\right)\Bigg|_{Y=Y^{\rm on-shell}}\,.
\ee

To find the one point function of the boundary operator, $\langle J_a\rangle$, we consider the variation of the effective action as
\bear
\delta W(X)&=&\delta\left(Y^T F X+{1\over 2} Y^T M Y\right)\Bigg|_{Y=Y^{\rm on-shell}}\nonumber\\&=&\delta Y^T\left(FX+MY\right)\bigg|_{Y=Y^{\rm on-shell}}+(Y^{\rm on-shell})^T F\delta X\,,
\eear
where the first term vanishes due to the equation of motion satisfied by $Y^{\rm on-shell}$, so we conclude that
\be
\langle J_a\rangle={\delta W(X)\over \delta X^a}=\left((Y^{\rm on-shell})^T F\right)_a= \left(Y^{\rm on-shell}\right)^I F_{Ia}\,.
\ee
Since $Y^{\rm on-shell}$ is the one-point function profile of the bulk dynamical modes, the above relation gives a relation between the bulk one-point function $Y^{\rm on-shell}$, and the boundary one-point function $\langle J_a\rangle$. This precisely corresponds to the bulk to boundary relation between one-point functions as in (\ref{onep1}) and (\ref{onep2}) which is illustrated in Figure \ref{fig2x}(a).
Therefore, one can identify $F$ with our previous projection operator $\cal P$: multiplying the matrix $F$ maps to the action of projection operator $\cal P$,
\be
F_{Ia}\longleftrightarrow {\cal P}\,.\label{FtoP}
\ee
Roughly speaking, the bulk index $I$ maps to the holographic radial coordinate $r$ in this correspondence.

To find the two-point functions $G_{ab}$ from $W(X)$ by (\ref{onetwo}), let us compute the on-shell action with the on-shell solution of $Y$ (\ref{onshell}),
\be
W(X)=S(X,Y^{\rm on-shell})=\left(Y^T F X+{1\over 2} Y^T M Y\right)\Bigg|_{Y=Y^{\rm on-shell}}=-{1\over 2}X^T F^T M^{-1} F X\,,
\ee
from which one obtains
\be
G_{ab}=(-i) {\delta^2 W(X)\over\delta X^a\delta X^b}=i \left(F^T M^{-1} F\right)_{ab}=i  F_{aI}\left(M^{-1}\right)^{IJ} F_{Jb}\,.\label{Gab}
\ee
Now, here comes the reward of our discussion: the bulk path integral of dynamical variables $Y$ with Dirichlet boundary condition $X=0$,
\be
\int {\cal D}Y\,\, e^{iS(X=0,Y)}=\int {\cal D} Y\,\,\exp\left({i\over 2} Y^T M Y\right)\,,
\ee
defines a quantum field theory by itself in the bulk, and the corresponding bulk-to-bulk two point function is given by
\be
g^{IJ}\equiv \langle Y^I Y^J\rangle_{\rm bulk}\equiv {\int {\cal D} Y\,\,Y^I Y^J \exp\left({i\over 2} Y^T M Y\right)\over \int {\cal D} Y\,\,\exp\left({i\over 2} Y^T M Y\right)}=i\left(M^{-1}\right)^{IJ}\,.\label{gIJ}
\ee
Comparing (\ref{Gab}) for $G_{ab}$ with (\ref{gIJ}) for $g^{IJ}$, we find that they are simply related by multiplying the matrix $F$ to each bulk indices $I$ and $J$.
Since multiplying $F$ corresponds to acting the projection operator $\cal P$  on the holographic coordinate as seen in (\ref{FtoP}), this explains the bulk to boundary relation between the two-point functions (\ref{projection}) as illustrated in Figure \ref{fig2x}(b),
\be
G(x,x')={\cal P}|_r {\cal P}|_{r'} g(x,r|x',r')\,.
\ee

Finally, let us define the boundary-to-bulk propagator ${\cal G}^I_a$ obtained by multiplying $F$ to one bulk index of $g^{IJ}$, or equivalently by acting the projection operator $\cal P$ on one radial coordinate of the bulk-to-bulk propagator $g^{IJ}=i\left(M^{-1}\right)^{IJ}$ as in (\ref{boundarytobulk}),
\be
{\cal G}^I_a\equiv i \left(M^{-1}\right)^{IJ} F_{Ja}\,.
\ee
Looking back the expression of the solution of the bulk equation of motion (\ref{onshell}),
\be
\left(Y^{\rm on-shell}\right)^I=-\left(M^{-1}\right)^{IJ} F_{Jb} X^b\,,
\ee
we observe that ${\cal G}^I_a$ viewed in terms of the bulk index $I$ is a solution of the bulk equation of motion with a UV boundary condition $X^b=(-i)\delta^{ab}$.
This explains (\ref{bulkeom2}) and the UV boundary condition (\ref{UVboundary}) in the property 3) of the previous section.

\section{Gradient correction to photon emission rate at strong coupling\label{mainsec}}

Based on the framework of real-time perturbation theory in AdS/CFT correspondence presented in the preceding section, we will compute in this section the correction to the photon emission rate at linear order in the shear component of the velocity gradient $\sigma_{ij}$.

To obtain the corrections to the photon emission rate at linear order in the shear mode of velocity gradients $\sigma_{ij}$, one needs to compute the corrections to the real-time correlation functions of the field theory U(1) currents, $G^<_{\mu\nu}(k)$, which can be obtained from the corrections to $G^{(rr)}_{\nu\mu}(k)$, $G^{(ra)}_{\nu\mu}(k)$ and $G^{(ar)}_{\nu\mu}(k)$ via (\ref{grrraar}),
\be
G^<_{\mu\nu}(k)=G^{(rr)}_{\nu\mu}(k)-{1\over 2}G^{(ra)}_{\nu\mu}(k)+{1\over 2}G^{(ar)}_{\nu\mu}(k)\,.
\ee
These in turn can be computed from the corrections to (the Wigner transform of ) the bulk-to-bulk propagators of the bulk U(1) gauge field, $g^{(\alpha\beta)}_{\mu\nu}(x,r|x',r')$, and applying the projection operations
with respect to $r$ and $r'$.

Since we are interested only in the linear term in $\sigma_{ij}$ which is already first order in derivative, we can safely assume that $\sigma_{ij}$ is space-time homogeneous (that is, constant) for our purpose of identifying the first correction in derivative expansion: the contributions coming from the space-time variations of $\sigma_{ij}$ should be considered as second or higher order corrections.
Because of this assumption, the resulting correction to the real-time correlation functions $g^{(\alpha\beta)}_{\mu\nu}(x,r|x',r')$ proportional to $\sigma_{ij}$ will be translationally invariant, depending only on $(x-x')$, and the Wigner transforms can be simply replaced by the Fourier transforms.

The $g^{(\alpha\beta)}_{\mu\nu}(x,r|x',r')$ are the bulk two point functions of the U(1) Maxwell gauge field, so the only way they are affected by $\sigma_{ij}$ is through the background metric perturbation induced by having the velocity gradient of $\sigma_{ij}$ in a near equilibrium black-hole geometry, corresponding to hydrodynamic evolution of the gauge theory plasma. This perturbation was computed in the method of fluid-gravity correspondence \cite{Bhattacharyya:2008jc}.
Explicitly, the bulk metric has the form \footnote{It should have been written in Eddington-Finkelstein (EF) coordinate, but for constant $\sigma_{ij}$ we are assuming the change from EF to the present Schwarz coordinate does not affect the form of $\delta g_{ij}$.}
\be
ds^2={dr^2\over f(r,T)r^2}+r^2\left(-f(r,T)u_\mu u_\nu dx^\mu dx^\nu+\left(\eta_{\mu\nu}-u_\mu u_\nu\right)dx^\mu dx^\nu\right)+\delta g_{\mu\nu} dx^\mu dx^\nu\,,\label{origmet}
\ee
where
\be
f(r,T)=1-\left(\pi T\over r\right)^4\,,
\ee
and the local temperature $T$ and the fluid velocity $u_\mu$ are now slowly varying in space-time, representing hydrodynamic evolution of the gauge theory plasma close to equilibrium. In our case, we only turn on $\sigma_{ij}$. Due to their variations, the Einstein equation necessitates the existence of the correction $\delta g_{\mu\nu}$ which can be systematically computed in gradient expansion of $T$ and $u_\mu$.
From $\delta g_{\mu\nu}$ one can compute the viscous corrections to the energy-momentum tensor due to slow variations of $T$ and $u_\mu$ in derivative expansion. For our purpose, we only need the correction proportional to $\sigma_{ij}$ in the local rest frame where $u^\mu=(1,\vec 0)$, and only the traceless component of $\delta g_{ij}$ is induced by this. It is given by
\be
\delta g_{ij}(r)=S(r) \sigma_{ij}\,,\label{metricpert}
\ee
where $S(r)$ can be read off, for example, in Refs.\cite{Bhattacharyya:2008jc,Torabian:2009qk} as
\be
S(r)={r^2\over 2}{1\over\pi T}\left(\pi-2\arctan\left(r\over\pi T\right)+\log\left[\left(1+\left(\pi T\over r\right)\right)^2 \left(1+\left(\pi T\over r\right)^2\right)\right]\right)\,.
\ee

\begin{figure}[t]
	\centering
	\includegraphics[width=8cm]{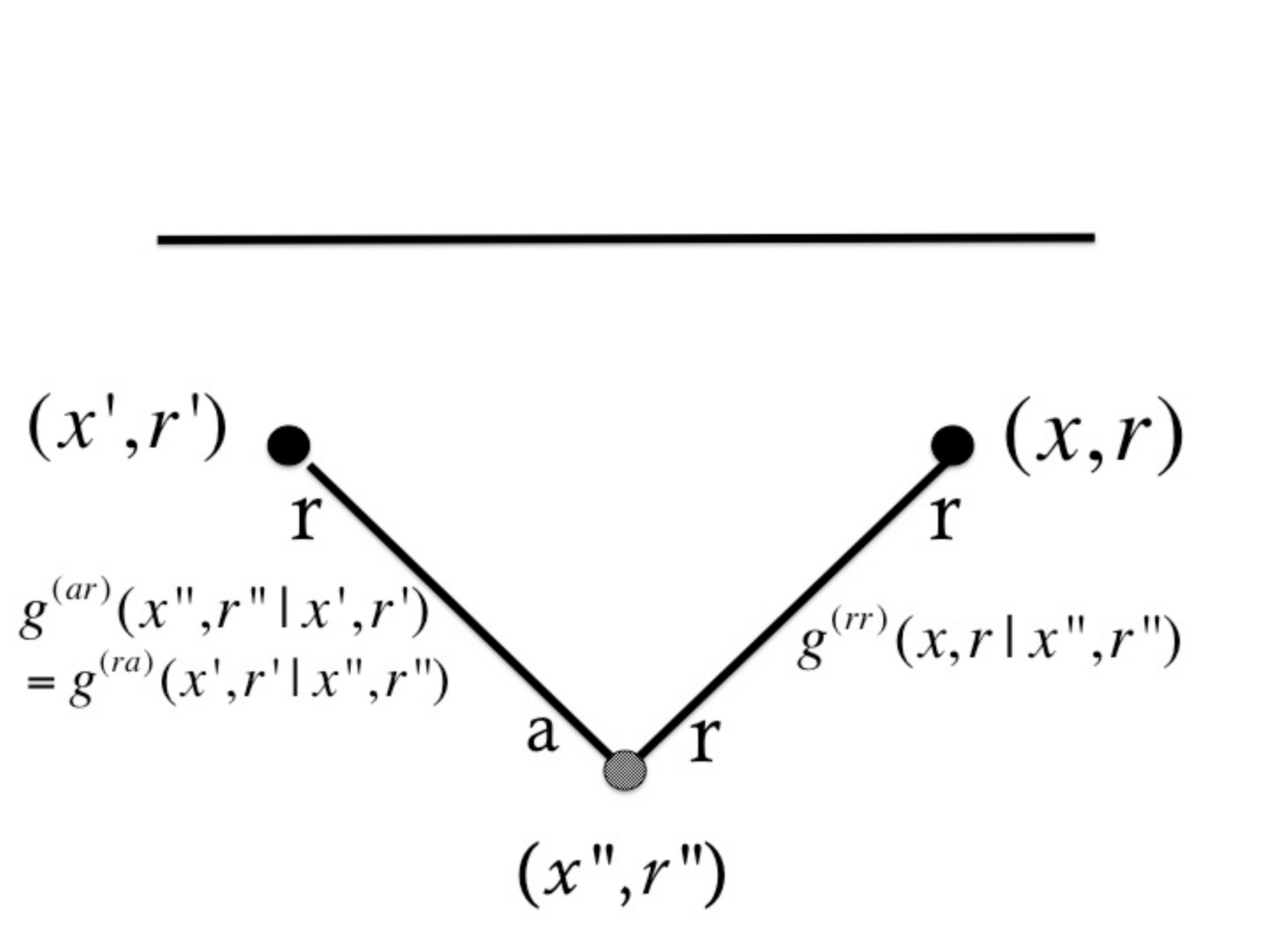}\includegraphics[width=8cm]{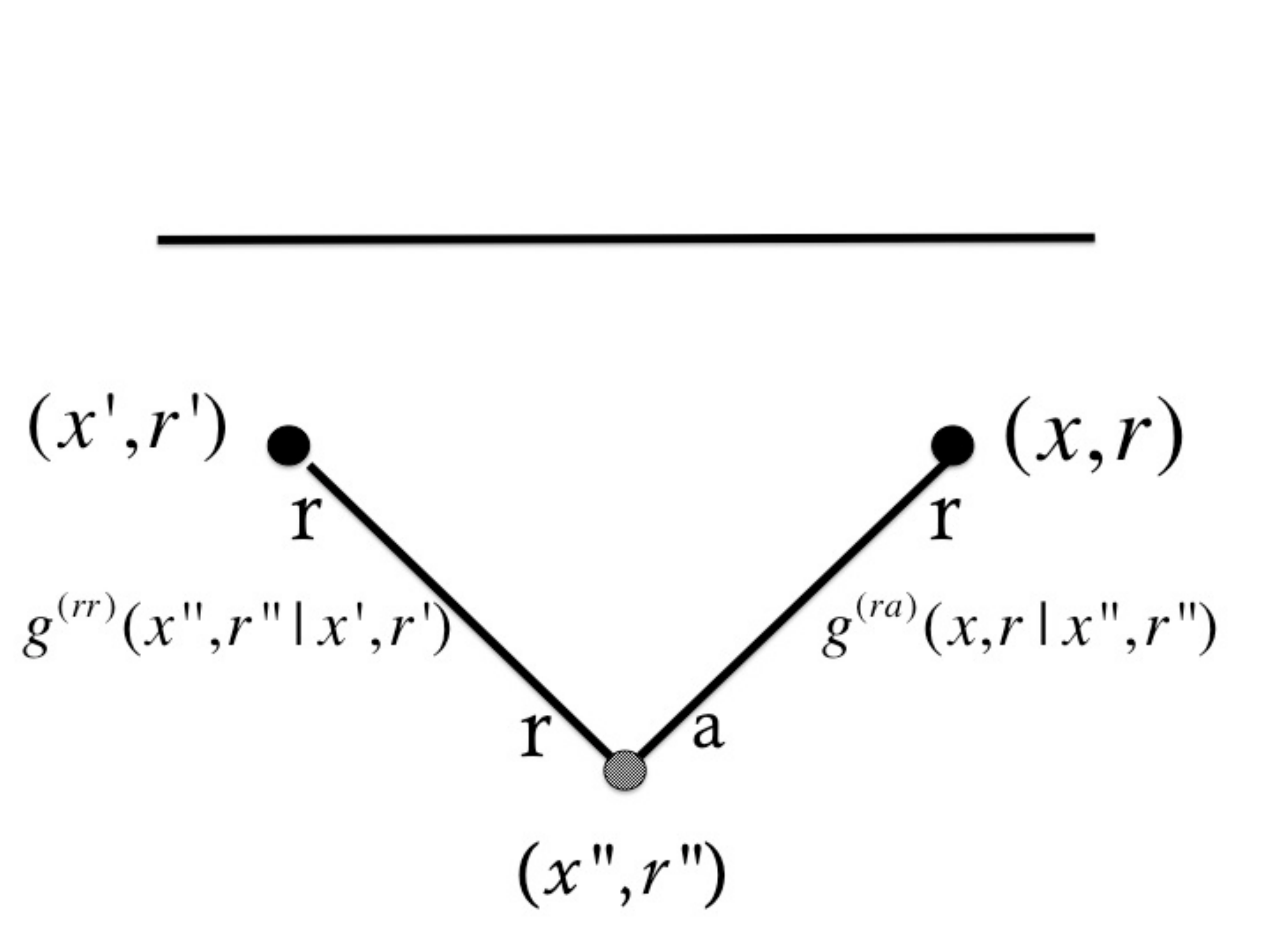}
		\caption{Real-time Feynman diagrams for $\delta g^{(rr)}_{\mu\nu}(x,r|x',r')$. There are two diagrams. \label{fig4}}
\end{figure}
The metric perturbation (\ref{metricpert}) proportional to $\sigma_{ij}$ should be considered as an external perturbation to the bulk U(1) gauge field dynamics, so in terms of Schwinger-Keldysh variables, it should correspond to
\be
\delta g^{(1)}_{ij}=\delta g^{(2)}_{ij}=\delta g_{ij}\,,
\ee
or equivalently,
\be
\delta g^{(r)}_{ij}=\delta g_{ij}\,,\quad \delta g^{(a)}_{ij}=0\,.
\ee
This gives rise to the following change of the bulk Schwinger-Keldysh action of the U(1) gauge field in (ra) variables,
\bear
\delta {S}_{\rm SK}=\delta { S}_1-\delta {S}_2&=&{1\over 32\pi G_5}\sigma_{ij}
\int d^5 x \sqrt{-g_5}\left({S(r)\over r^4}F^{(a)}_{iN}F^{(r)}_{jQ}g^{NQ}\right)\nonumber\\&=&
{1\over 32\pi G_5}\sigma_{ij}
\int d^4x dr \left({S(r)\over r}F^{(a)}_{iN}F^{(r)}_{jQ}g^{NQ}\right)\,,\label{bulkvertex}
\eear
where the metric appearing in the above is the unperturbed black-hole metric in the rest frame.
Note that the induced vertex has one ``r'' variable and one ``a'' variable. It is now straightforward to compute the corrections to the various real-time two point functions at linear order in the above action perturbation in the framework of real-time perturbation theory in the bulk\footnote{The metric (\ref{origmet}) also has a correction from the varying fluid velocity $u_i=\sigma_{ij}x^j$ itself near the local rest frame $x=0$, which is linear in $x$. This induces a non-local contribution to the current correlation functions via Wigner transform near $x=0$: since we would have a Fourier transform involving linear term in $x^j$, the result is suppressed by additional power of $T/k=T/\omega$ compared to the direct local contribution coming from $\delta g_{ij}$ we are computing. In weak coupling quasi-particle picture, this non-local contribution can be attributed to the effects from non-locality of finite Compton wavelength $\Delta x\sim 1/k$ of quasi-particles, which has been neglected for hard photons $\omega > T$ compared to the local effects coming from disturbances of particle distribution functions. The $\delta g_{ij}$ we are considering is a holographic strong coupling analogue of the latter effects.} .

\begin{figure}[t]
	\centering
	\includegraphics[width=8cm]{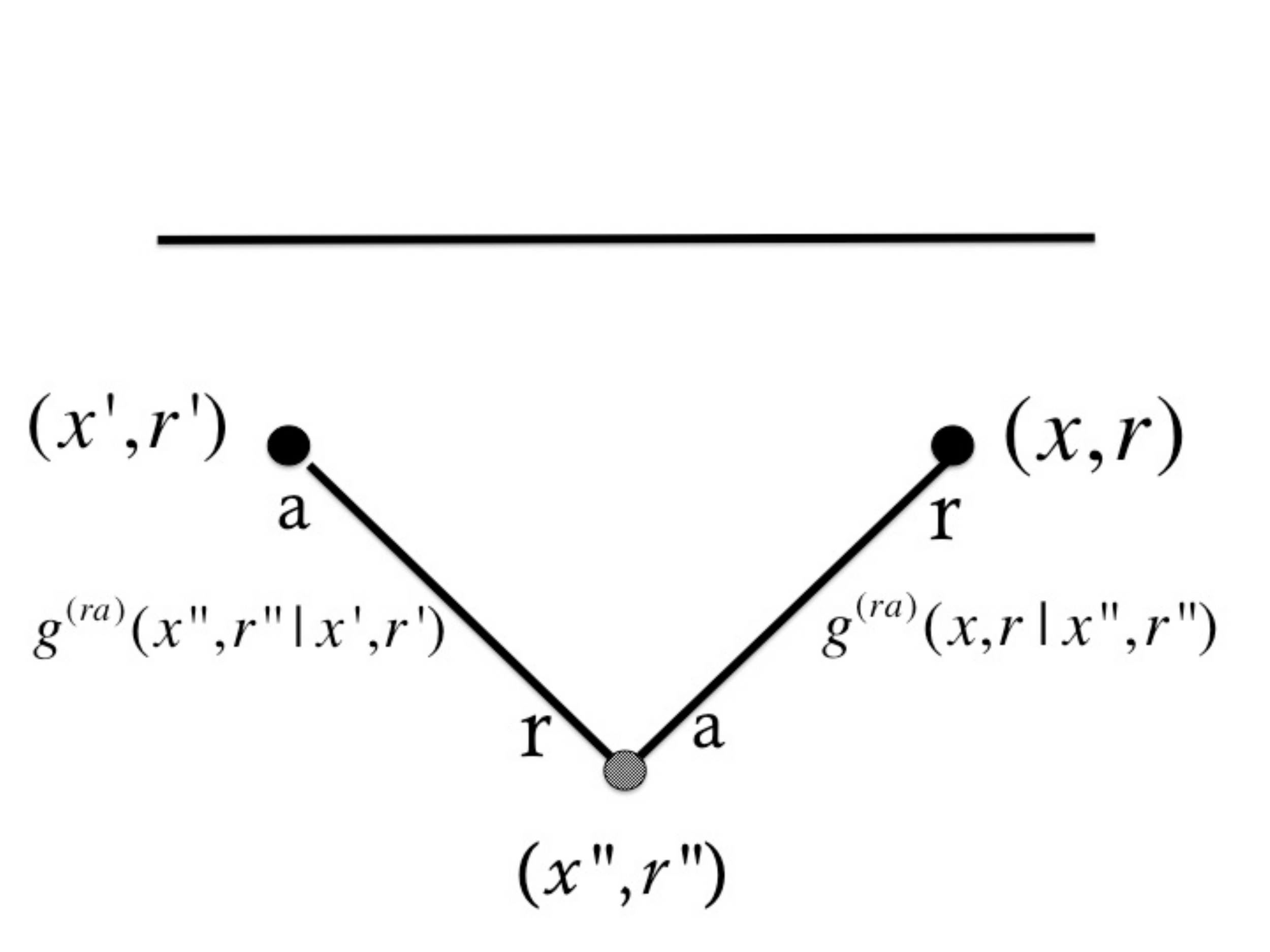}\includegraphics[width=8cm]{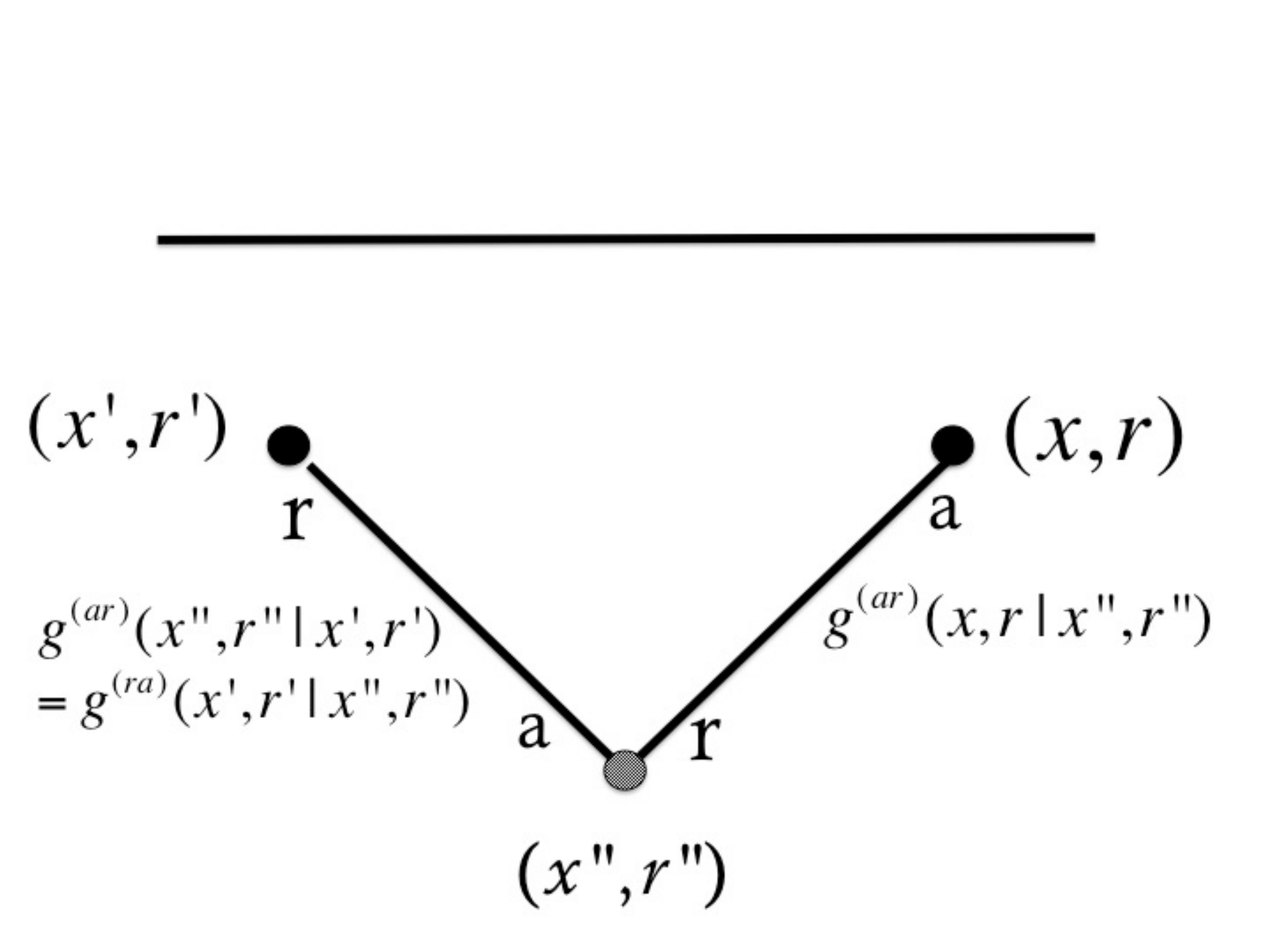}
		\caption{Real-time Feynman diagrams for $\delta g^{(ra)}_{\mu\nu}(x,r|x',r')$ (left) and $\delta g^{(ar)}_{\mu\nu}(x,r|x',r')$ (right).\label{fig5}}
\end{figure}

Figure \ref{fig4} depicts the (space-time) Feynman diagrams for the contributions to $\delta g^{(rr)}_{\mu\nu}$: there are two possible diagrams depending on how (ra) indices are contracted. The lines represent the zero'th order equilibrium bulk-to-bulk propagators, and the blob at $(x'',r'')$ is the vertex insertion induced by (\ref{bulkvertex}) proportional to $\sigma_{ij}$.  Figure \ref{fig5} shows the diagrams for $\delta g^{(ra)}_{\mu\nu}$ and $\delta g^{(ar)}_{\mu\nu}$, where only one type of diagram is possible for each, due to the fact that $g^{(aa)}_{\mu\nu}$ does not exist.

Note that all diagrams involve at least one bulk-to-bulk propagator $g^{(ra)}(x,r|x'',r'')$ or $g^{(ra)}(x',r'|x'',r'')$. Recalling that the retarded propagator $g^R$ is simply $(-i)$ times of $g^{(ra)}$,
and since the region beyond event horizon located at $r=r_H=\pi T$ (that is, the region $r<r_H$) is causally disconnected to the region of $r>r_H$ which contains the UV boundary,
the retarded bulk-to-bulk propagators $g^{(ra)}(x,r|x'',r'')$ or $g^{(ra)}(x',r'|x'',r'')$, which respect the causal structure of the geometry, will simply vanish if $r''<r_H$. Therefore, the integration over $r''$ of the position of the perturbation vertex will automatically be reduced to the region above the horizon $r>r_H$.

\begin{figure}[t]
	\centering
	\includegraphics[width=8cm]{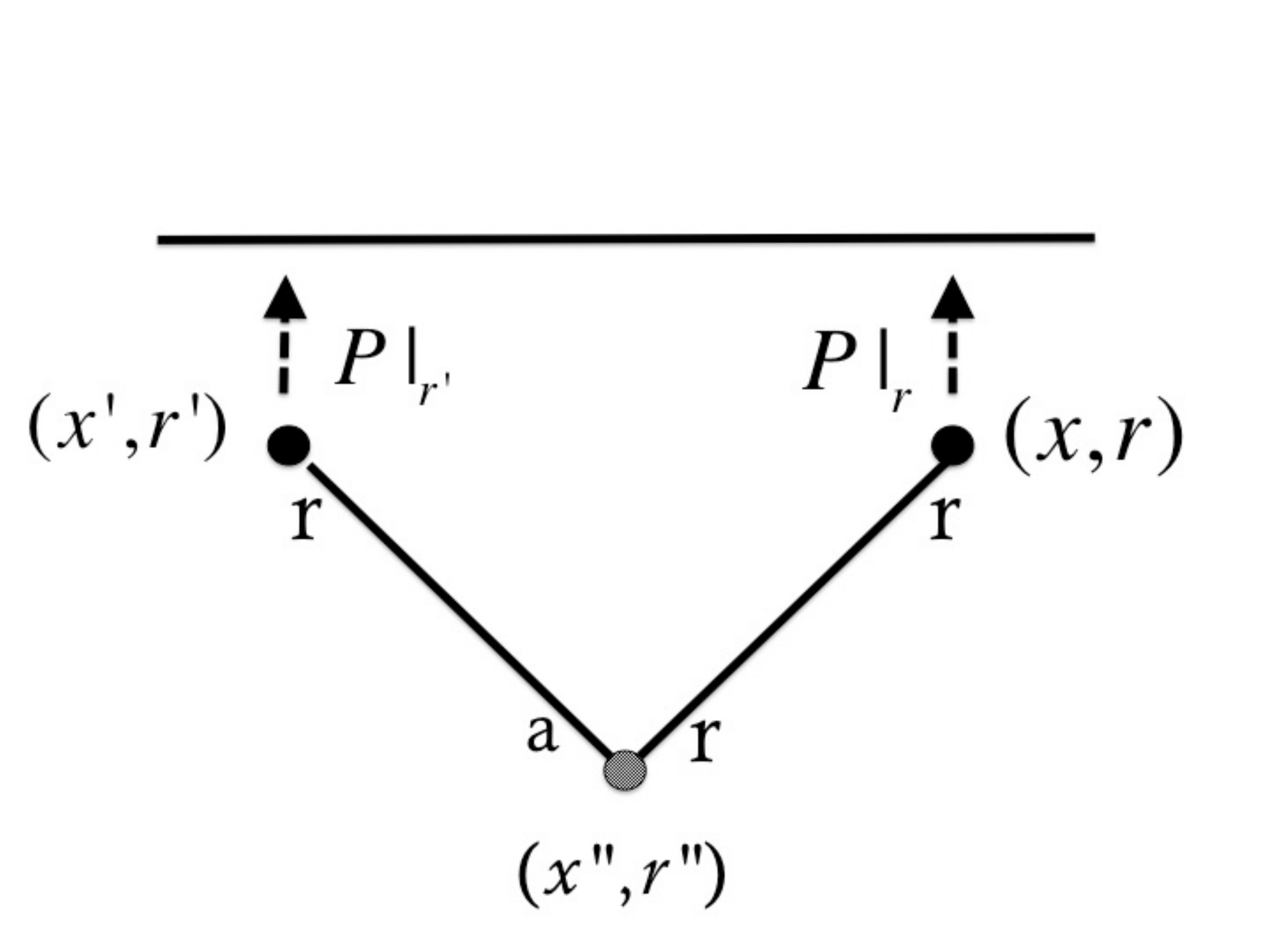}\includegraphics[width=8cm]{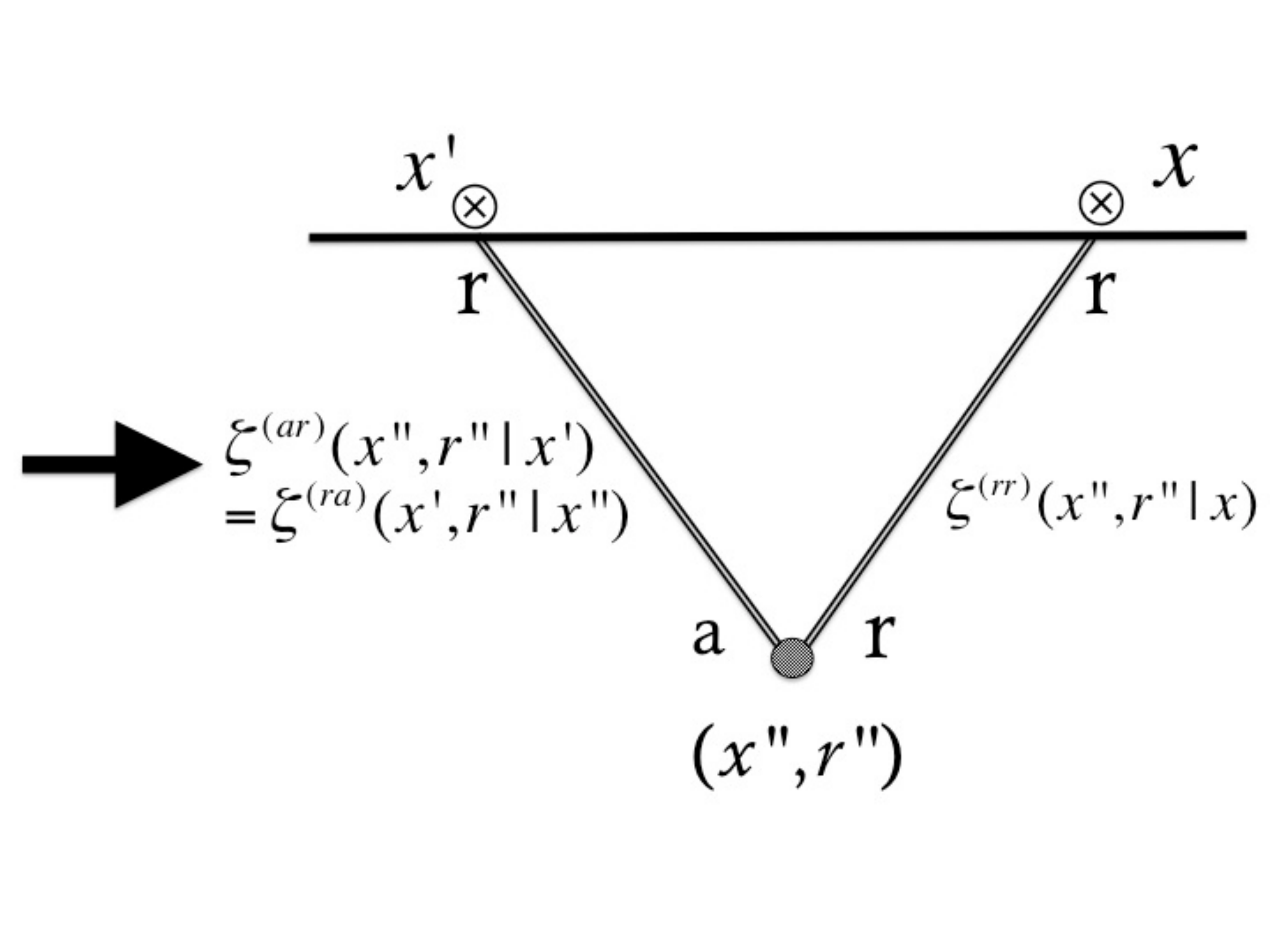}
		\caption{One of the two real-time Feynman diagrams for $\delta G^{(rr)}_{\mu\nu}(x|x')$ (right) obtained from $\delta g^{(rr)}_{\mu\nu}(x,r|x',r')$ by projection operators $\cal P$ (left). The other diagrams are similarly constructed from Figure \ref{fig4} and Figure \ref{fig5}. \label{fig6}}
\end{figure}

At the end, what we would like to compute is the corrections to the field theory two point functions $G^{(\alpha\beta)}_{\mu\nu}$, which are obtained from the above bulk-to-bulk two point functions by the boundary projection ${\cal P}$. Therefore, one can choose to apply the projection $\cal P$ to each diagram to get diagrammatic contributions to the field theory two-point functions directly. Applying the projection ${\cal P}$ simply replaces the lines of bulk-to-bulk equilibrium propagators with the boundary-to-bulk equilibrium propagators ${\cal G}^{(\alpha\beta)}_{\mu\nu}$, so that the resulting diagram in Figure \ref{fig6} (right) looks almost same as before, except that the new lines now represent the boundary-to-bulk propagators ${\cal G}^{(\alpha\beta)}_{\mu\nu}$. We will try to compute these diagrams with vector indices contracted with transverse photon polarization vector, $\epsilon^\mu$, in the following.

Because $\sigma_{ij}$ is homogeneous in $x^\mu$ up to our order of interest, we can work in the Fourier space with respect to 4 dimensional coordinates $x^\mu$, while keeping the radial direction $r$ unchanged, by introducing
\be
{\cal G}^{(\alpha\beta)}_{\mu\nu}(x,r|x')=\int {d^4 k\over (2\pi)^4} \,e^{ik(x-x')} {\cal G}^{(\alpha\beta)}_{\mu\nu}(k,r)\,,
\ee
et cetera. The relation (\ref{ravsar2}) between ${\cal G}^{(ra)}$ and ${\cal G}^{(ar)}$ is translated to
\be
{\cal G}^{(ra)}_{\mu\nu}(k,r)={\cal G}^{(ar)}_{\nu\mu}(-k,r)\,,\label{relation3}
\ee
which will be used in the following.
Working out the Feynman rules of the induced vertex (\ref{bulkvertex}) in momentum space, it is straightforward to write down the resulting corrections to the two point functions in momentum space: for example, the $\delta G^{(ra)}_{\mu\nu}(k)$ is given by
\bear
\delta G^{(ra)}_{\mu\nu}(k)&=&{i\over 32\pi G_5}\sigma_{ij}\int_{r_H}^\infty dr\,{S(r)\over r}\Bigg[
\left(\partial_r {\cal G}^{(ra)}_{i\nu}(k,r)\right)\left(\partial_r {\cal G}^{(ra)}_{\mu j}(k,r)\right)g^{rr}\label{finalgra}\\
&+& \left(k_i {\cal G}^{(ra)}_{\rho\nu}(k,r)-k_{\rho}{\cal G}^{(ra)}_{i\nu}(k,r)\right)\left(k_j {\cal G}^{(ra)}_{\mu\sigma}(k,r)-k_{\sigma}{\cal G}^{(ra)}_{\mu j}(k,r)\right)g^{\rho\sigma}\Bigg]\,,\nonumber
\eear
where $\rho,\sigma$ run only over the 4 dimensional coordinates, and we have used (\ref{relation3}) in arriving to the above.
The expressions for $\delta G^{(ar)}_{\mu\nu}(k)$ and $\delta G^{(rr)}_{\mu\nu}(k)$
look almost identical to the above with the replacements of ${\cal G}^{(ra)}$ with appropriate ${\cal G}^{(\alpha\beta)}$ corresponding to the Feynman diagrams of Figure \ref{fig4} and Figure \ref{fig5} (after the projection operators are applied).
We would like to contract the results with the photon polarization tensors to have $\delta G^{(\alpha\beta)}_{\mu\nu}(k)(\epsilon^\mu)^*\epsilon^\nu\equiv \delta G^{(\alpha\beta)}(k;\epsilon)$.

Fortunately, once we compute $\delta G^{(ra)}(k;\epsilon)$, we don't need to compute other $\delta G^{(ar)}(k;\epsilon)$ and $\delta G^{(rr)}(k;\epsilon)$: they can be obtained from $\delta G^{(ra)}(k;\epsilon)$ by
\be
\delta G^{(ar)}(k;\epsilon)=\delta G^{(ra)}(-k;\epsilon)=-\left[\delta G^{(ra)}(k;\epsilon)\right]^*\,,\label{FDT1}
\ee
and
\be
\delta G^{(rr)}(k;\epsilon)=(1+2n_B(\omega)){\rm Re}\left[\delta G^{(ra)}(k;\epsilon)\right]=-(1+2n_B(\omega)){\rm Im}\left[\delta G^{R}(k;\epsilon)\right]\,,\label{FDT2}
\ee
where $n_B(\omega)$, $\omega\equiv k^0$, is the Bose-Einstein distribution with the equilibrium temperature $T$, and the retarded correlation function $G^R$ is defined to be $-i G^{(ra)}$. From the relation (\ref{grrraar}),
\be
G^<_{\mu\nu}= G^{(rr)}_{\nu\mu}-{1\over 2}G^{(ra)}_{\nu\mu}+{1\over 2}G^{(ar)}_{\nu\mu}\,,\label{grrraar2}
\ee
this means that the desired correction $\delta G^<(k;\epsilon)$ entering the photon emission rate is simply given in terms of $\delta G^{(ra)}(k;\epsilon)$ by
\be
\delta G^<(k;\epsilon)=2n_B(\omega){\rm Re}\left[\delta G^{(ra)}(k;\epsilon)\right]=-2n_B(\omega) {\rm Im}\left[ \delta G^R(k;\epsilon)\right]\,.\label{fin1}
\ee

To show (\ref{FDT1}) and (\ref{FDT2}), first observe that the equilibrium two point functions entering the Feynman diagrams of Figure \ref{fig6} satisfy the equilibrium fluctuation-dissipation relation~\footnote{Strictly speaking, we need this relation only for transverse components because of the contraction with the photon polarization at the end. In fact, this relation is unambiguously valid without worrying about gauge fixing only for the transverse components of the two point functions. Therefore, we will be loose in writing down this relation for all components as in (\ref{FDT3}).}
\be
{\cal G}^{(rr)}_{\mu\nu}(k,r)=(1+2n_B){\rm Re}\left[{\cal G}^{(ra)}_{\mu\nu}(k,r)\right]=-(1+2n_B){\rm Im}\left[{\cal G}^{R}_{\mu\nu}(k,r)\right]\,.\label{FDT3}
\ee
Recall also the relation (\ref{relation3}),
\be
{\cal G}^{(ra)}_{\mu\nu}(k,r)={\cal G}^{(ar)}_{\nu\mu}(-k,r)\,.\label{FDT4}
\ee
Finally, since the retarded two point function ${\cal G}^R_{\mu\nu}(x,r|x')\equiv (-i) {\cal G}^{(ra)}_{\mu\nu}(x,r|x')$ is real valued, its Fourier transform satisfies the reality condition
\be
{\cal G}^R_{\mu\nu}(k,r)=\left[{\cal G}^R_{\mu\nu}(-k,r)\right]^*\,,
\ee
which in turn gives
\be
{\cal G}^{(ra)}_{\mu\nu}(k,r)=-\left[{\cal G}^{(ra)}_{\mu\nu}(-k,r)\right]^*\,.\label{FDT5}
\ee
The (\ref{FDT3}), (\ref{FDT4}), and (\ref{FDT5}) allow us to replace ${\cal G}^{(rr)}_{\mu\nu}(k,r)$ by
\be
{\cal G}^{(rr)}_{\mu\nu}(k,r)={1\over 2}(1+2n_B)\left({\cal G}^{(ra)}_{\mu\nu}(k,r)-{\cal G}^{(ar)}_{\nu\mu}(k,r)\right)\,.\label{FDT6}
\ee
Using the relation (\ref{FDT6}) to replace ${\cal G}^{(rr)}_{\mu\nu}(k,r)$ in the two Feynman diagrams for $\delta {G}^{(rr)}_{\mu\nu}(k)$ in Figure \ref{fig4} (after the projection operators are applied), one can easily check that
the resulting sum of the two diagrams is precisely equal to
\be
{1\over 2}(1+2n_B)\left(\delta { G}^{(ra)}_{\mu\nu}(k)-\delta { G}^{(ar)}_{\nu\mu}(k)\right)\,.\label{FDT7}
\ee
Since the relations (\ref{FDT4}) and (\ref{FDT5}) between the retarded and advanced functions are generally valid, we also have
\be
\delta {G}^{(ra)}_{\mu\nu}(k)=\delta { G}^{(ar)}_{\nu\mu}(-k)=-\left[\delta{G}^{(ra)}_{\mu\nu}(-k)\right]^*\,.\label{FDT8}
\ee
If one wishes, one can check them from the Feynman diagrams for $\delta { G}^{(ra)}_{\mu\nu}(k,r)$ and $\delta { G}^{(ar)}_{\mu\nu}(k,r)$ in Figure \ref{fig5} (after the projection operators are applied) using the relations (\ref{FDT4}) and (\ref{FDT5}). The results (\ref{FDT7}) and (\ref{FDT8}) finally prove the relations (\ref{FDT1}) and (\ref{FDT2}).

The relation (\ref{FDT2}) means that the fluctuation-dissipation relation persists to hold even for our first order gradient corrections to the two point functions, with the same $(1+2n_B)$ factor given by the zero'th order equilibrium temperature. This seems in line with the discussion in Ref.\cite{Mukhopadhyay:2012hv}. This is {\it a priori} not obvious since the first order corrections are out-of-equilibrium characteristics of the system. We expect the fluctuation-dissipation relation to be violated in higher order corrections than the first order.

The expression (\ref{fin1}) we have derived,
\be
\delta G^<(k;\epsilon)=2n_B(\omega){\rm Re}\left[\delta G^{(ra)}(k;\epsilon)\right]\,,\label{fin2}
\ee
then requires us to compute only $\delta G^{(ra)}(k;\epsilon)$, which is given by the expression (\ref{finalgra}) contracted with the photon polarization vector. After contracting with the polarization, what appears in the expression is the polarization contracted retarded boundary-to-bulk two point function, either $(\epsilon^\mu)^* {\cal G}^{(ra)}_{\mu\nu}(k,r)$ or $\epsilon^\nu {\cal G}^{(ra)}_{\mu\nu}(k,r)$. Since $\epsilon^\mu$ is transverse, these objects should satisfy the transverse part of the bulk U(1) Maxwell equation (see (\ref{bulkeom2})) which is easily derived to be
\be
\left[\partial_r\left(r^3 f(r)\partial_r \right)+{1\over r}\left({\omega^2\over f(r)}-|\vec k|^2\right) \right](\epsilon^\mu)^* {\cal G}^{(ra)}_{\mu\nu}(k,r)=0\,.\label{treom}
\ee
The UV boundary condition at $r\to\infty$ given by (\ref{UVboundary}) is translated for the Fourier transforms as
\be
\lim_{r\to\infty} (\epsilon^\mu)^* {\cal G}^{(ra)}_{\mu\nu}(k,r)=-i(\epsilon_\nu)^*\,,
\ee
and finally one has to impose the incoming boundary condition at the horizon $r=r_H$ which is the correct boundary condition for the retarded two point function. These conditions uniquely determine $(\epsilon^\mu)^* {\cal G}^{(ra)}_{\mu\nu}(k,r)$. In our case of photons, we only need to consider the light-like on-shell momenta with $|\vec k|=\omega$, and an analytic form of the solution is available in this case (first found in Ref.\cite{CaronHuot:2006te}) as
\bear
(\epsilon^\mu)^* {\cal G}^{(ra)}_{\mu\nu}(k,r)&=&-i(\epsilon_\nu)^* \left(1-\left(\pi T\over r\right)^2\right)^{-i{\omega\over 4\pi T}}\left(1+\left(\pi T\over r\right)^2\right)^{-{\omega\over 2\pi T}}\label{form1}\\&\times& { _2F_1\left(1-{1\over 2}(1+i){\omega\over 2\pi T},-{1\over 2}(1+i){\omega\over 2\pi T};1-i{\omega\over 2\pi T};{1\over 2}\left(1-\left(\pi T\over r\right)^2\right)\right)
\over _2F_1\left(1-{1\over 2}(1+i){\omega\over 2\pi T},-{1\over 2}(1+i){\omega\over 2\pi T};1-i{\omega\over 2\pi T};{1\over 2}\right)}\,,\nonumber
\eear
in terms of the hypergeometric function $_2F_1(a,b;c;z)$. Similarly, we have
\bear
\epsilon^\nu {\cal G}^{(ra)}_{\mu\nu}(k,r)&=&-i \epsilon_\mu \left(1-\left(\pi T\over r\right)^2\right)^{-i{\omega\over 4\pi T}}\left(1+\left(\pi T\over r\right)^2\right)^{-{\omega\over 2\pi T}}\label{form2}\\&\times& { _2F_1\left(1-{1\over 2}(1+i){\omega\over 2\pi T},-{1\over 2}(1+i){\omega\over 2\pi T};1-i{\omega\over 2\pi T};{1\over 2}\left(1-\left(\pi T\over r\right)^2\right)\right)
\over _2F_1\left(1-{1\over 2}(1+i){\omega\over 2\pi T},-{1\over 2}(1+i){\omega\over 2\pi T};1-i{\omega\over 2\pi T};{1\over 2}\right)}\,.\nonumber
\eear
With these formulae, $\delta G^{(ra)}(k;\epsilon)$ from (\ref{finalgra}) can now be expressed as a radial integral of an analytic expression involving hypergeometric functions. We will further simplify the expression shortly using the equation of motion, but before doing that we need to discuss the necessary infrared regularization on the horizon.

A short inspection shows that the $r$ integral for $\delta G^{(ra)}(k;\epsilon)$ from (\ref{finalgra})
with the above hypergeometric functions contains the terms which behave near the horizon as
\be
\sim (\#) \sigma_{ij}\epsilon_i (\epsilon_j)^* \omega^2 \int_{r_H} dr \,(r-r_H)^{-i{\omega\over 2\pi T}-1}\sim \lim_{r\to r_H} (r-r_H)^{-i{\omega\over 2\pi T}}\,,\label{IR}
\ee
which is ill-defined (though {\it not} divergent). Specifically, these terms are present in the first piece of (\ref{finalgra}) and the $\rho=\sigma=t$ component of the second piece, with the same strength so that they add up together. The origin of this behavior is entirely due to the factor
\be
\left(1-\left(\pi T\over r\right)^2\right)^{-i{\omega\over 4\pi T}}\,,
\ee
in front of the above retarded two point functions at light-like momenta, which is universal for any retarded propagator with incoming boundary condition at the horizon without depending on specific details of a bulk theory.
 The presence of this behavior near the horizon should map to an infrared problem at light-like momenta in the corresponding dual field theory side.
To tame this infrared behavior, we use the well-known technique of shifting
\be
\omega\to \omega+i\epsilon\,,\quad \epsilon=0^+\,,
\ee
in the retarded two point function ${\cal G}^{(ra)}(\omega)$: it is clear that this prescription cures the ambiguity in (\ref{IR}) by giving an extra factor of $(r-r_H)^\epsilon$, making the horizon limit well-defined. The final regularized integral has a good well-defined limit in $\epsilon\to 0^+$. The physics meaning of this shift is to give an asymptotic damping to the retarded two point function, such that in a large time limit, $G^{(ra)}(t)$ vanishes as {\it at least} $e^{-\epsilon t}$ or faster in $t\to+\infty$ limit. Recall that in the Fourier expansion of the retarded function
\be
G^{(ra)}(t)=\int {d\omega\over (2\pi)}\, e^{-i\omega t} G^{(ra)}(\omega)\,,
\ee
the $G^{(ra)}(\omega)$ is analytic in the upper half plane of complex $\omega$, ensuring that $G^{(ra)}(t)=0$ for $t<0$. There may be poles or branch cuts in the lower half planes giving rise to non-zero $G^{(ra)}(t)$ for $t>0$. The imaginary parts of the poles or branch cuts give the damping rate: $e^{-i\omega_p t}\sim e^{{\rm Im}(\omega_p) t}$, where $\omega_p$ is the pole or branch cut location with ${\rm Im}(\omega_p)\le 0$. The shift $\omega\to\omega+i\epsilon$ in the argument of $G^{(ra)}(\omega)$ shifts the imaginary parts of the poles or branch cuts by an amount $\omega_p\to \omega_p-i\epsilon$, enhancing the damping rate by $\epsilon$. The analyticity in the upper half plane is intact.

The expression for $\delta G^{(ra)}(k;\epsilon)$ with the above analytic formulae for $(\epsilon^\mu)^* {\cal G}^{(ra)}_{\mu\nu}(k,r)$ and $\epsilon^\nu {\cal G}^{(ra)}_{\mu\nu}(k,r)$ can be written after some algebra as
\bear
\delta G^{(ra)}(k;\epsilon)&=&-{i\over 32\pi G_5}{\sigma_{ij}\over C^2}\int_{r_H}^\infty dr\,{S(r)\over r}\Bigg[\epsilon_i(\epsilon_j)^*\left((\partial_r H(r))^2 r^2 f(r)-{H(r)^2\over r^2}\left({\omega^2\over f(r)}-|\vec k|^2\right)\right)\nonumber\\&+&k_i k_j {H(r)^2\over r^2}\Bigg]\,,\label{semifinal}
\eear
where in the middle of computation, we have used the transversality $\epsilon^i k_i=0$, and
\bear
H(r)&\equiv&
\left(1-\left(\pi T\over r\right)^2\right)^{-i{\omega\over 4\pi T}}\left(1+\left(\pi T\over r\right)^2\right)^{-{\omega\over 2\pi T}}\\&\times& _2F_1\left(1-{1\over 2}(1+i){\omega\over 2\pi T},-{1\over 2}(1+i){\omega\over 2\pi T};1-i{\omega\over 2\pi T};{1\over 2}\left(1-\left(\pi T\over r\right)^2\right)\right)\,,\nonumber
\eear
and
\be
C\equiv   {_2F_1\left(1-{1\over 2}(1+i){\omega\over 2\pi T},-{1\over 2}(1+i){\omega\over 2\pi T};1-i{\omega\over 2\pi T};{1\over 2}\right)}\,.
\ee
In all the above expressions, we assume the infrared regularization of $\omega\to\omega+i0^+$ introduced before.
The above result (\ref{semifinal}) can be simplified further by using the equation of motion satisfied by $H(r)$: recall that $H(r)$ is a solution of the transverse part of the U(1) Maxwell equation (\ref{treom}),
\be
\partial_r\left(r^3 f(r)\partial_r H(r)\right)+{1\over r}\left({\omega^2\over f(r)}-|\vec k|^2\right) H(r)=0\,.\label{treom2}
\ee
Using this and performing integration by part the first term in (\ref{semifinal}), one finally arrives at our compact result,
\bear
\delta G^{(ra)}(k;\epsilon)={i\over 32\pi G_5}{\sigma_{ij}\over C^2}\int_{r_H}^\infty dr\,\Bigg[\epsilon_i(\epsilon_j)^*\partial_r\left({S(r)\over r^2}\right)r^3 f(r)H(r)\partial_r H(r)-k_i k_j {S(r)H(r)^2\over r^3}\Bigg]\,,\nonumber\\ \label{final}
\eear
where we have used the fact that the boundary term from the integration by part,
\be
-{i\over 32\pi G_5}{\sigma_{ij}\over C^2}\epsilon_i (\epsilon_j)^*\left({S(r)\over r^2}r^3 f(r)H(r)\partial_r H(r)\right)\Bigg|^\infty_{r_H}\,,
\ee
vanishes both at $r=\infty$ and $r=r_H$: we emphasize that the vanishing at the horizon occurs precisely with our infrared regularization $\omega\to\omega+i0^+$, and it wouldn't happen without it. The (\ref{final}) with (\ref{fin2}) can now give the expression for the correction to the photon emission rate with polarization $\epsilon^\mu$ as
\be
{d \Gamma^{\rm shear}\over d^3 \vec k}(\epsilon^\mu)={e^2\over (2\pi)^3 2\omega}\epsilon^\mu (\epsilon^\nu)^* \delta G^<_{\mu\nu}(k)={e^2\over (2\pi)^3 2\omega} 2n_B(\omega){\rm Re}\left[\delta G^{(ra)}(k;\epsilon)\right]\,.
\ee

To finally find the correction to the total emission rate, summing over photon polarization vectors replaces
\be
\epsilon_i (\epsilon_j)^*\to \delta_{ij}-\hat k_i\hat k_j\,,
\ee
in the above, and since $\sigma_{ij}$ is traceless, the $\delta_{ij}$ term does not contribute, so that the result becomes
\bear
\sum_{\epsilon^\mu}\delta G^{(ra)}(k;\epsilon)=-{i N_c^2\over 16\pi^2 C^2}\hat k^i\hat k^j\sigma_{ij}\int_{r_H}^\infty dr\,\Bigg[\partial_r\left({S(r)\over r^2}\right)r^3 f(r)H(r)\partial_r H(r)+\omega^2 {S(r)H(r)^2\over r^3}\Bigg]\,,\nonumber\\ \label{final2}
\eear
where we have used $G_5=\pi/(2N_c^2)$. Using (\ref{fin2}), the final expression for $\Gamma^{(1)}(\omega)$ in the total rate,
\be
{d\Gamma^{\rm shear}\over d^3 \vec k}={e^2 \over T}\Gamma^{(1)}(\omega)\hat k^i\hat k^j \sigma_{ij}\,,
\label{correction2}
\ee
is given by
\bear
\Gamma^{(1)}(\omega)&=&{1\over (2\pi)^3 2\omega} 2n_B(\omega){\frac{N_{c}^2T}{16\pi^2}}\\ &\times& {\rm Im}\left[{1\over C^2}\int_{r_H}^\infty dr\,\Bigg[\partial_r\left({S(r)\over r^2}\right)r^3 f(r)H(r)\partial_r H(r)+\omega^2 {S(r)H(r)^2\over r^3}\Bigg]\right]\,.\nonumber
\eear
If one changes the integration variable to
\be
u\equiv \left(\pi T\over r\right)^2\,,
\ee
the expression becomes the one given in the introduction (\ref{intro1}).
\begin{figure}[t]
	\centering
	\includegraphics[width=10cm]{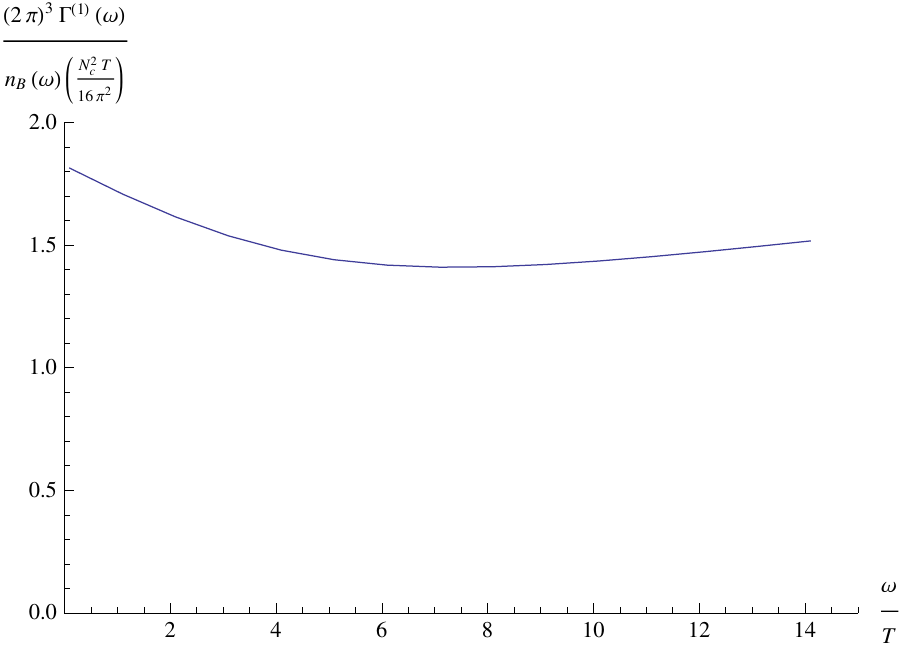}
		\caption{The plot of $\Gamma^{(1)}(\omega)/n_B(\omega)$ in unit of $\sigma/(2\pi)^3$, $\sigma\equiv  {\frac{N_{c}^2T}{16\pi^2}}$ being the electric conductivity. \label{fig9}}
\end{figure}
In Figure \ref{fig9}, we show the plot of $\Gamma^{(1)}(\omega)/n_B(\omega)$ as a function of $\omega/T$. The values in unit of $\sigma/(2\pi)^3$, $\sigma\equiv  {\frac{N_{c}^2T}{16\pi^2}}$ being the electric conductivity, look quite constant over a large range of $\omega/T$, ranging from 1.4 to 1.8.

\section{Discussion}

\begin{figure}[t]
	\centering
	\includegraphics[width=10cm]{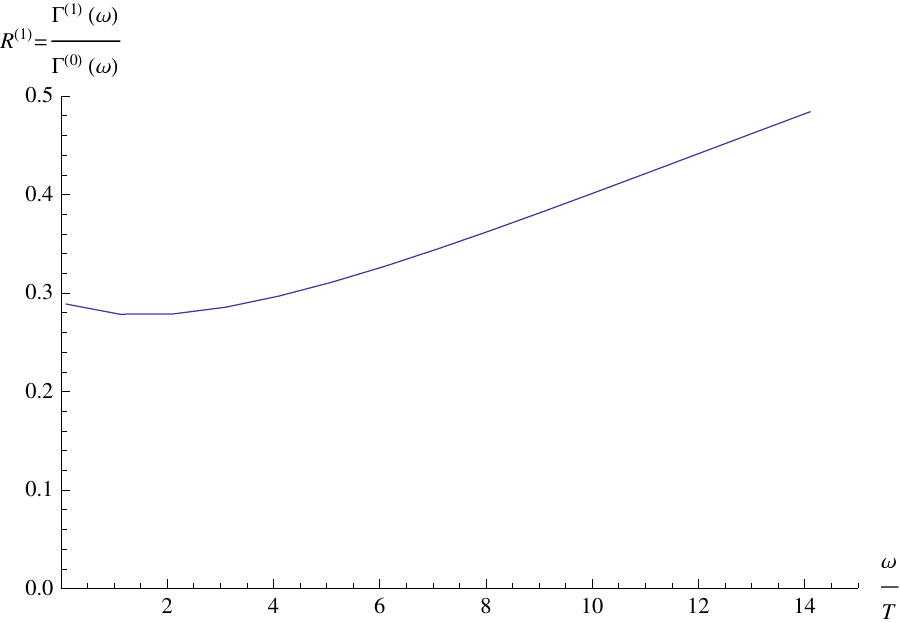}
		\caption{The plot of $R^{(1)}=\Gamma^{(1)}(\omega)/\Gamma^{(0)}(\omega)$ as a function of $\omega/T$.\label{fig10}}
\end{figure}
To draw a meaningful conclusion to realistic QCD from our results, it is perhaps useful to consider the dimensionless ratio of the gradient correction to the zero'th order emission rate,
\be
R^{(1)}\equiv{\Gamma^{(1)}(\omega)\over \Gamma^{(0)}(\omega)}\,,
\ee
where the $\Gamma^{(0)}(\omega)$ is the equilibrium photon emission rate in the derivative expansion
 \be
 {d\Gamma\over d^3 \vec k}=e^2\Gamma^{(0)}(\omega)+{e^2 \over T}\Gamma^{(1)}(\omega)\hat k^i\hat k^j \sigma_{ij}+\cdots\,,
\ee
which was first computed in Ref.\cite{CaronHuot:2006te},
\bear
\Gamma^{(0)}(\omega)&=&{1\over (2\pi)^3 2\omega} \epsilon^\mu(\epsilon^\nu)^* G^<_{\mu\nu}(k)\Bigg|_{\omega=|\vec k|}    \times 2\\
&=& {1\over (2\pi)^3 \omega}\,\,2n_B(\omega){\rm Re}\left[ \epsilon^\mu(\epsilon^\nu)^* G^{(ra)}_{\mu\nu}(k)\right]
={1\over (2\pi)^3 \omega}\,\,2n_B(\omega){\rm Im}\left[{1\over C}\,{\cal P}|_r\cdot H(r)\right]\nonumber\\
&=&{1\over (2\pi)^3 \omega}\,\,n_B(\omega) {N_c^2 T \omega\over 32\pi}\left|{_2F_1}\left(1-{1\over 2}(1+i){\omega\over 2\pi T},1+{1\over 2}(1-i){\omega\over 2\pi T}; 1-i{\omega\over 2\pi T};-1\right)\right|^{-2}\,,\nonumber
\eear
where the last factor 2 in the first line comes from the polarization summation, and $H(r)$ and $C$ are as defined above. In the second line, we have used the fact that
\be
\epsilon^\mu(\epsilon^\nu)^* {\cal G}^{(ra)}_{\mu\nu}(k,r)=-{i\over C} H(r)\,,
\ee
which can be seen from (\ref{form1}) and (\ref{form2}), as well as the relation $G={\cal P}|_r\cdot {\cal G}$. To get to the last line, we followed the same steps in Ref.\cite{CaronHuot:2006te}, using $G_5=\pi/(2N_c^2)$. In Figure \ref{fig10}, we show the plot of $R^{(1)}$ as a function of $\omega/T$. The value starts at around 0.3 and keeps increasing for higher frequency. We see that the gradient correction may not be negligible, and our result provides a useful strong coupling benchmark for this correction.

We discuss phenomenological importance of the gradient correction we find in the on-going relativistic heavy-ion experiments.
The initial quark-gluon plasma right after collision has no fluid velocity gradient, and as the fluid velocity develops via pressure gradient toward radially outward direction, the velocity near the boundary of the plasma can reach to a fractional order one value of the speed of light, while the velocity at the center of the plasma stays small. Since the typical size of the plasma is about 10 fm, the velocity gradient is roughly $\sigma_{ij}\sim 1/(10 \,{\rm fm})\sim 20$ MeV. Taking the temperature of about 200 MeV, we have $\sigma_{ij}/T\sim 0.1$. As the value of $R^{(1)}\lesssim 0.5$, the relative size of the gradient correction to the equilibrium rate is about $5\%$ which seems somewhat small, but not negligible. We think this indicates a good convergence of the gradient expansion scheme. However, our estimate for $\sigma_{ij}$ may be too crude, if we consider fluctuations as well as flows, and the size of $10\%$ or greater is not unexpected. It would be interesting to implement our result in the existing numerical simulations of heavy-ion collisions to quantify
its effects.

As a final remark, our computational technique should be applicable to other observables which can be obtained from real time correlation functions, such as jet quenching, quark diffusion, etc.

\vskip 1cm \centerline{\large \bf Acknowledgment} \vskip 0.5cm

We are indebted to Derek Teaney who introduced this problem to us. We thank Jean F. Paquet, Misha Stephanov, Derek Teaney, Chaolun Wu, and Yi Yin for helpful discussions and comments.

 \vfil

\end{document}